\documentclass[12pt]{article}
\usepackage[reqno]{amsmath}
\usepackage{mathrsfs}
\usepackage{amssymb}
\usepackage{bbm}
\usepackage{epsfig}
\usepackage{array}
\usepackage{float}
\usepackage{color}
\usepackage{graphicx}

\usepackage{a4}

\usepackage{a4wide}
\usepackage{wasysym}

\def\gs{\mathrel{
   \rlap{\raise 0.511ex \hbox{$>$}}{\lower 0.511ex \hbox{$\sim$}}}}
\def\ls{\mathrel{
   \rlap{\raise 0.511ex \hbox{$<$}}{\lower 0.511ex \hbox{$\sim$}}}}
\newcommand{\obb}{0\mbox{$\nu\beta\beta$}}
\newcommand{\onbb}{neutrino-less double beta decay}
\newcommand{\ba}{\begin{array}{c}}
\newcommand{\baz}{\begin{array}{cc}}
\newcommand{\bad}{\begin{array}{ccc}}
\newcommand{\bea}{\begin{equation} \begin{array}{c}}
\newcommand{\eea}{ \end{array} \end{equation}}
\newcommand{\ea}{\end{array}}
\newcommand{\D}{\displaystyle}
\newcommand{\dms}{\mbox{$\Delta m^2_{\odot}$}}
\newcommand{\dma}{\mbox{$\Delta m^2_{\rm A}$}}
\newcommand{\meff}{\mbox{$\left| m_{ee} \right|$}}
\newcommand{\eV}{\mbox{ eV}}

\newcommand{\be}{\begin{eqnarray}}
\newcommand{\ee}{\end{eqnarray}}

\newcommand{\sss}{\sin^2 \theta_{12}}
\newcommand{\sch}{\sin^2 \theta_{13}}

\begin{document}

\begin{titlepage}

\title{
\vspace*{-2.0cm}
\hfill {\small TUM-HEP--614/05}\\[-5mm]
\hfill {\small hep--ph/0512143}\\[20mm]
\bf\large
Improved Limit on $\theta_{13}$ and Implications for 
Neutrino Masses in Neutrino-less Double Beta Decay and 
Cosmology\\[5mm]\ }

\author{Manfred Lindner\thanks{email: \tt lindner@ph.tum.de}~~,~~
Alexander Merle\thanks{email: \tt alexander$\_$merle@ph.tum.de}~~,~~
Werner Rodejohann\thanks{email: \tt werner$\_$rodejohann@ph.tum.de} 
\\ \\
{\normalsize \it Physik-Department, Technische Universit\"at M\"unchen,}\\
{\normalsize \it  James-Franck-Strasse, D-85748 Garching, Germany}
}
\date{}
\maketitle
\thispagestyle{empty}

\begin{abstract}
\noindent
We analyze the impact of a measurement, or of an improved bound, on 
$\theta_{13}$ for the determination of the effective neutrino 
mass in neutrino-less double beta decay and cosmology. In particular, 
we discuss how an improved limit on (or a specific value of) $\theta_{13}$ 
can influence the determination of the neutrino mass spectrum via 
neutrino-less double beta decay. We also discuss the interplay with 
improved cosmological neutrino mass searches.
\end{abstract}

\end{titlepage}


\section{\label{sec:intro}Introduction}

The absolute mass scale and the Majorana nature of neutrinos are 
among the central topics of the future research program in neutrino 
physics \cite{APSgen,APSmass}. In addition, the value of the currently 
unknown mixing matrix element $|U_{e3}| = \sin \theta_{13}$ is of 
central importance, since it is a strong discriminator for neutrino 
mass models. The magnitude of $|U_{e3}|$ is also important for 
future efforts to probe leptonic $CP$ violation and/or the mass 
ordering in oscillation experiments (see e.g.~\cite{Huber:2004ug}). 
Neutrino-less double beta decay (\obb) is the best known method to 
address both the Majorana nature of neutrinos, as well as the absolute 
mass scale. Several ongoing and planned experiments, such as 
NEMO3 \cite{NEMO}, CUORICINO \cite{CUORICINO}, CUORE \cite{CUORE}, 
MAJORANA \cite{MAJORANA}, GERDA \cite{GERDA}, EXO \cite{EXO}, 
MOON \cite{MOON}, COBRA \cite{COBRA}, XMASS, DCBA \cite{DCBA}, 
CANDLES \cite{CANDLES}, CAMEO \cite{CAMEO} 
aim at observing the process 
\[ (A,Z) \rightarrow (A,Z + 2) + 2 \, e^-~. \] 
If mediated by light Majorana neutrinos, the square root of the 
decay width of \obb\ is proportional to a so-called effective 
mass which is given by the following coherent sum: 
\be
\meff \equiv \left| \sum_i m_i \, U_{ei}^2 \right|~,
\label{eq:meff1}
\ee
where $m_i$ is the mass of the $i^{th}$ neutrino mass state
and where the sum is over all light neutrino mass states. 
$U_{ei}$ are the elements of the leptonic mixing matrix 
\cite{PMNS} which we parameterize here as 
\be 
\label{eq:Upara}
U = \left( \bad 
c_{12} c_{13} & s_{12} c_{13} & s_{13} \, e^{-i \delta}  \\[0.2cm] 
-s_{12} c_{23} - c_{12} s_{23} s_{13} e^{i \delta} 
& c_{12} c_{23} - s_{12} s_{23} s_{13} e^{i \delta} 
& s_{23} c_{13}  \\[0.2cm] 
s_{12} s_{23} - c_{12} c_{23} s_{13} e^{i \delta} & 
- c_{12} s_{23} - s_{12} c_{23} s_{13} e^{i \delta} 
& c_{23} c_{13}  \\ 
               \ea   \right) 
 {\rm diag}(1, e^{i \alpha}, e^{i \beta}) \, , 
\ee
where we have used the usual notations $c_{ij} = \cos\theta_{ij}$, 
$s_{ij} = \sin\theta_{ij}$. $\delta$ is the Dirac $CP$-violation 
phase, $\alpha$ and $\beta$ are the two Majorana $CP$-violation 
phases \cite{BHP80}. 
The best current limit on the effective mass is given by the 
Heidelberg-Moscow collaboration \cite{HM} 
\be \label{eq:current} 
\meff \le 0.35\, \zeta~{\rm eV}~, 
\ee
where $\zeta={\cal O}(1)$ indicates an uncertainty due to 
uncertainties in the calculation of the nuclear matrix elements
of \obb. Similar results were obtained by the IGEX collaboration 
\cite{IGEX}. The above mentioned experiments will improve the
current bound by one order of magnitude\footnote{Those experiments
will of course also test the claimed evidence \cite{contr} by
part of the Heidelberg-Moscow collaboration.}.
In terms of the neutrino mass matrix,  
\be
m_{\alpha \beta} = U_{\alpha i} \, m_i \, \delta_{ij}  \, U^T_{j \beta}~,
\ee
\meff\ is nothing but the $ee$ element in the basis where the 
charged lepton mass matrix is real and diagonal. Neutrino-less 
double beta decay therefore probes directly an element of the 
mass matrix, which is a unique feature, not possible in the quark 
sector. $\meff$ in Eq.~(\ref{eq:meff1}) depends on the 
oscillation parameters, the Majorana phases and the overall 
neutrino mass scale. 
This means that \meff{} depends on 7 
out of 9 parameters contained in the neutrino mass matrix. 
It depends also on the neutrino mass ordering, which can be 
normal or inverted. It is interesting that the effective mass 
is a function of all unknowns of neutrino physics except for the 
Dirac phase\footnote{Within the usual parameterization 
Eq.~(\ref{eq:Upara}), it appears as if the Dirac phase is 
contained in \meff. However, this phase can be eliminated 
by means of a re-definition of the Majorana mass state 
$m_3$.} and $\theta_{23}$. The effective mass is therefore a 
probe of the neutrino mass scale and interestingly also of 
$\theta_{13}$. We focus in this work on the dependence 
on $\theta_{13}$, where significant improvements are 
expected. 
The current limit $\sin^22\theta_{13}<0.2$ 
will be somewhat improved by the on-going or up-coming 
neutrino beam experiments MINOS \cite{MINOS} and ICARUS 
\cite{ICARUS} as well as OPERA \cite{OPERA}, respectively.
Further significant improvement by one order of magnitude
compared to the existing bound will come within about 5 
years from reactor experiments such as Double Chooz 
\cite{DCHOOZ}. A few years later, the next generation of
superbeam experiments, T2K \cite{T2K} and No$\nu$A \cite{NOVA},
will further improve the measurements or the bound of 
$\theta_{13}$.
The absolute neutrino mass scale will also be attacked by
improved measurements of the end-point spectrum of tritium 
decay \cite{KATRIN}. Furthermore, improved cosmological 
measurements will improve our knowledge on the absolute 
neutrino mass scale from the role of neutrinos as hot
dark matter in the cosmological structure formation 
\cite{steen}. Altogether one can safely expect that the current limits 
will improve at least by one order of magnitude.

It is therefore interesting to analyze the interplay of 
$\theta_{13}$ with the neutrino mass scale, the neutrino 
mass ordering and \obb. 
In Section~\ref{sec:mee} we  
discuss the general dependence of the effective mass as 
a function of the neutrino observables. In 
Section~\ref{sec:NH} we discuss then in detail the case 
of normal mass ordering. We show that a very stringent 
limit on the effective mass leads to a limited range of 
values of the smallest neutrino mass, which translates 
into a certain range of the sum of neutrino masses as 
measurable in cosmology. The dependence on $\theta_{13}$ 
of these values is stressed. Section~\ref{sec:IH} deals 
then with the inverted mass ordering, and in 
Section~\ref{sec:combined} we discuss how $\theta_{13}$ 
influences the possibility to distinguish between normal
and inverted mass ordering via \obb. The uncertainty 
stemming from the nuclear matrix element calculations is 
also taken into account. 
Finally, we conclude in Section~\ref{sec:concl}.

\section{\label{sec:mee}Properties of the Effective Mass: General Aspects}

In this and the next two Sections we will discuss 
in some detail the value of the effective mass in terms of the known 
and unknown neutrino parameters \cite{others,NHvsIH,SC,SPP}, 
for a recent review see \cite{STP_rev}. 

\begin{figure}[tb]
\begin{center}
\epsfig{file=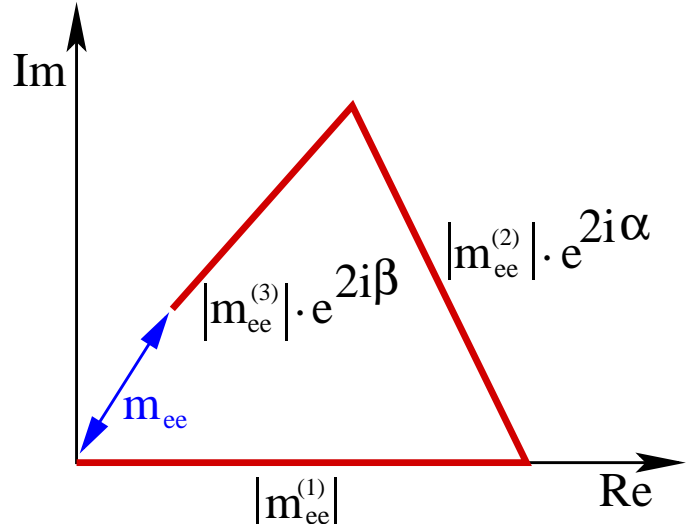,width=9cm}
\caption{\label{fig:vectors}The mass matrix element 
$m_{ee}$ as a sum of three complex vectors.}
\end{center}
\end{figure}

The effective mass is the absolute value of the mass matrix element 
$m_{ee}$, i.e., for three flavors it is a sum of three terms 
\be
\left| m_{ee} \right| \equiv \left| \sum U_{e i}^2 \, m_i \right| 
\mbox{ with } 
m_{ee} = |m_{ee}^{(1)}| + |m_{ee}^{(2)}| \, e^{2i\alpha} + 
|m_{ee}^{(3)}| \, e^{2i\beta} ~,
\label{eq:mee}
\ee
which is visualized in Fig.~\ref{fig:vectors} as the sum of three 
complex vectors $m_{ee}^{(1,2,3)}$. The Majorana phases $2\alpha$ and 
$2\beta$ correspond then to the relative orientation of the three vectors. 

In terms of the neutrino masses and mixing angles, we have 
\begin{eqnarray}
|m_{ee}^{(1)}| &=& m_1 \, |U_{e1}|^{2} = m_1 \, c_{12}^{2} \, c_{13}^{2} ~,
\nonumber\\
|m_{ee}^{(2)}| &=& m_2 \, |U_{e2}|^{2} = m_2 \, s_{12}^{2} \, c_{13}^{2} ~,\\
|m_{ee}^{(3)}| &=& m_3 \, |U_{e3}|^{2} = m_3 \, s_{13}^{2}~.\nonumber
\end{eqnarray}
Normal mass ordering corresponds to $m_3 > m_2 > m_1$, whereas for an 
inverted ordering we have $m_2 > m_1 > m_3$. The effective mass to be 
extracted from \onbb\ depends crucially on the neutrino mass spectrum. 
Fixing for the solar neutrino sector $\dms = m_2^2 - m_1^2 > 0$, we 
have for the atmospheric neutrino sector either $m_3^2 - m_1^2 > 0$ 
(normal ordering) or $m_3^2 - m_1^2 < 0$ (inverted ordering). We use 
a notation where $\dma \equiv |m_3^2 - m_1^2|$ is always positive.
The best-fit values and the $1\sigma$ and $3\sigma$ ranges of the 
oscillation parameters which will be used in this work are 
\cite{Schwetz} 
\begin{eqnarray}
\dms &=& 7.9^{+0.3\,, \,1.0}_{-0.3\,, \,0.8} \cdot 10^{-5} \eV^2~,\nonumber\\
\sss &=& 0.31^{+0.02\,, \,0.09}_{-0.03\,, \,0.07} ~,\nonumber\\
\dma &=&  2.2^{+0.37\,, \,1.1}_{-0.27\,, \,0.8} \cdot 10^{-3} \eV^2~,\\
\sin^2\theta_{23} &=& 0.50^{+0.06\,, \,0.18}_{-0.05\,, \,0.16} ~,\nonumber\\
\sch &<& 0.012~(0.046)~.\nonumber
\end{eqnarray}
The best-fit value for $\sch$ is 0. 
The two larger masses for each ordering are given in terms of 
the smallest mass and the mass squared differences as  
\be
\label{eq:masses}
\bad
\text{normal:}  &  m_2 = \sqrt{m_1^{2}+\dms} ~;~~~~~~~~~~~~~~~  m_3 = \sqrt{m_1^{2}+\dma} ~,\\
\text{inverted:}&  m_2 = \sqrt{m_3^{2}+\dms+\dma} ~;~~~~ m_1 = \sqrt{m_3^{2} + \dma} ~.
\ea
\ee

Of special interest are the following three extreme cases:  
\be 
\mbox{ normal hierarchy~(NH):} 
& ~~~~~~~~~~~~~ |m_3| \simeq \sqrt{\dma} \gg |m_{2}| \simeq \sqrt{\dms} \gg |m_1|~,\\[0.3cm]
\mbox{ inverted hierarchy~(IH):} 
& |m_2| \simeq |m_1| \simeq \sqrt{\dma} \gg |m_{3}| ~,\\[0.3cm]
\mbox{ quasi-degeneracy~(QD):} 
& ~~~~~~~~ m_0 \equiv |m_1| \simeq |m_2| \simeq |m_3|  \gg \sqrt{\dma} ~.
\label{eq:mass}
\ee
The order of magnitude of the effective mass in those spectra is 
$\sqrt{\dms}, \sqrt{\dma}$ and $m_0$, respectively 
(for recent analyzes of the effective mass in terms of the neutrino 
mass spectrum, see \cite{SC,SPP}). 
Within our parameterization Eq.~(\ref{eq:Upara}), 
it is sufficient to vary the Majorana phases $\alpha$ and $\beta$ 
between 0 and $\pi$ in order to obtain the full physical range of \meff. 
If there were processes sensitive to the off-diagonal 
elements of the neutrino mass matrix (from all that we know, 
there are not \cite{DL2}), then one would have to vary the 
phases in their full range between 0 and $2\pi$ to obtain the full 
physical range.

\begin{figure}[tb]
\begin{center}
\epsfig{file=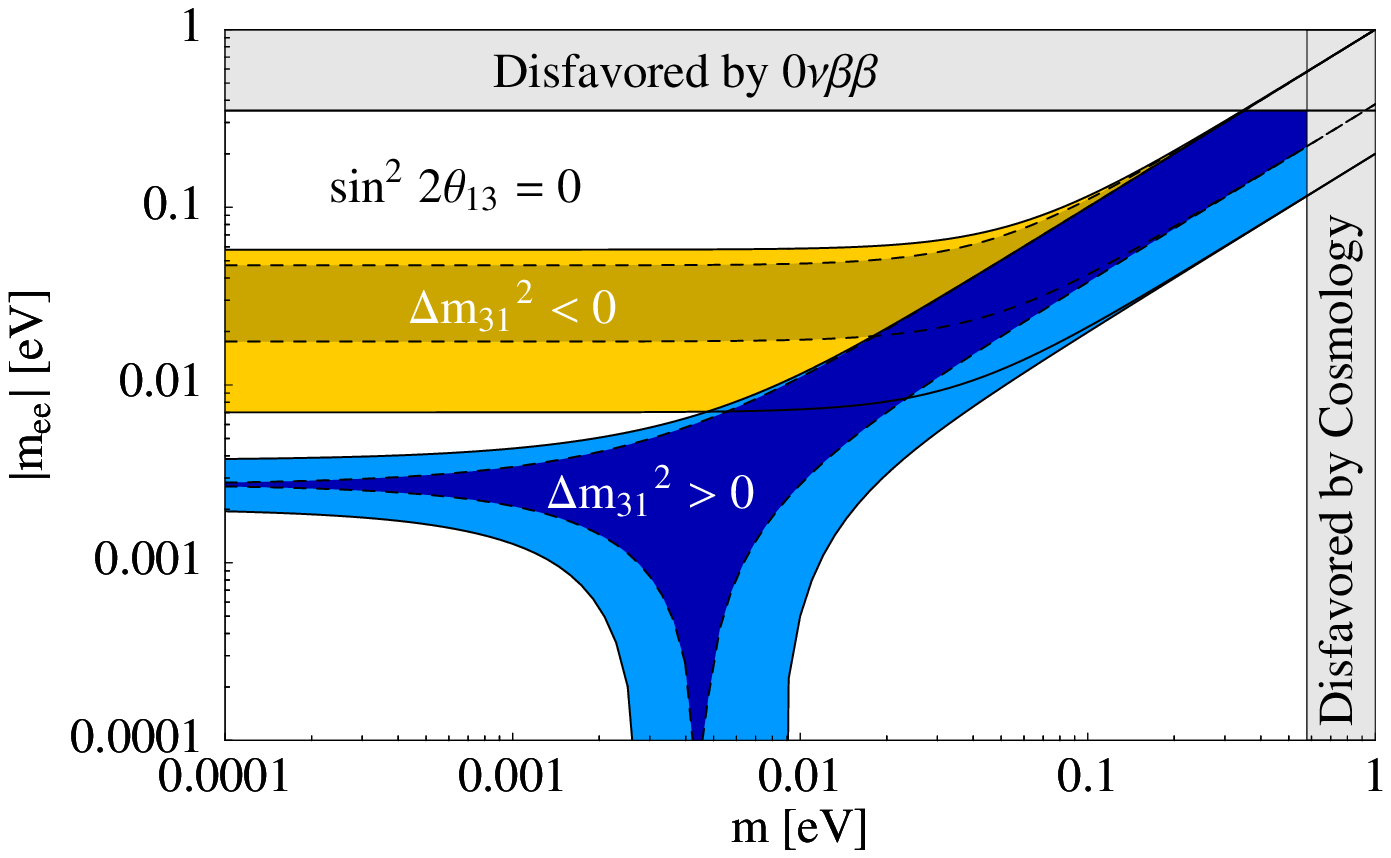,width=8cm,height=6cm}
\epsfig{file=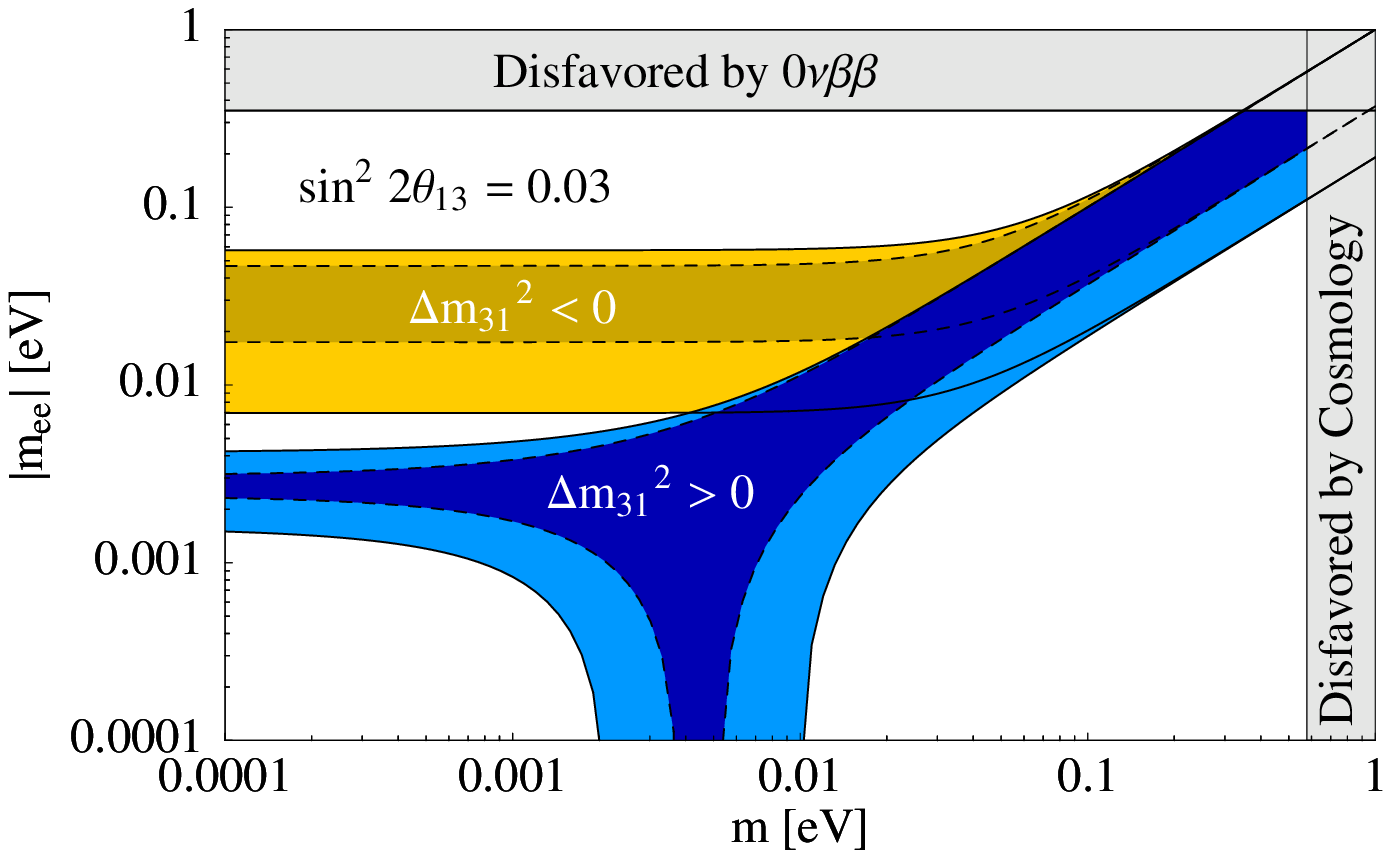,width=8cm,height=6cm}
\epsfig{file=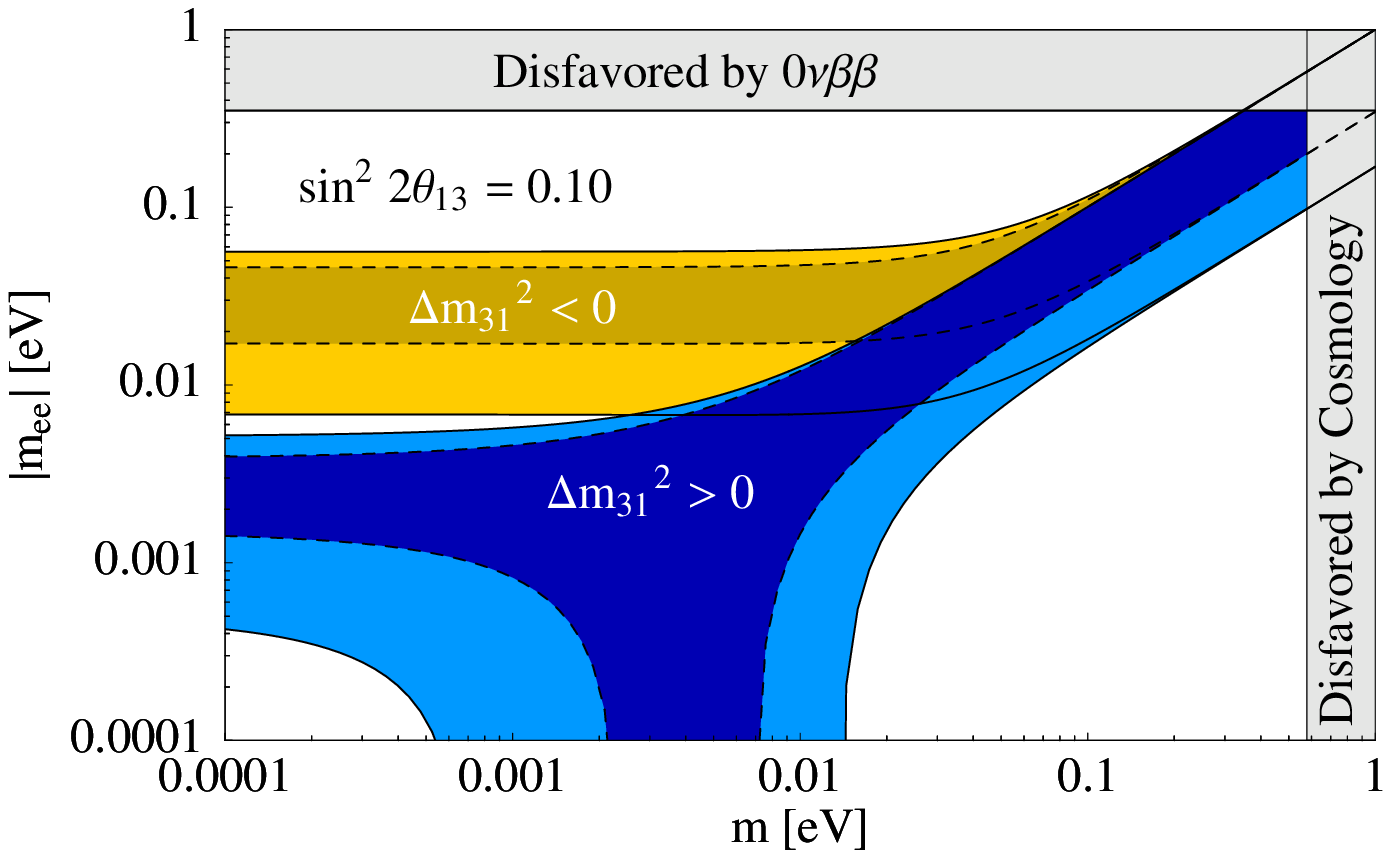,width=8cm,height=6cm}
\epsfig{file=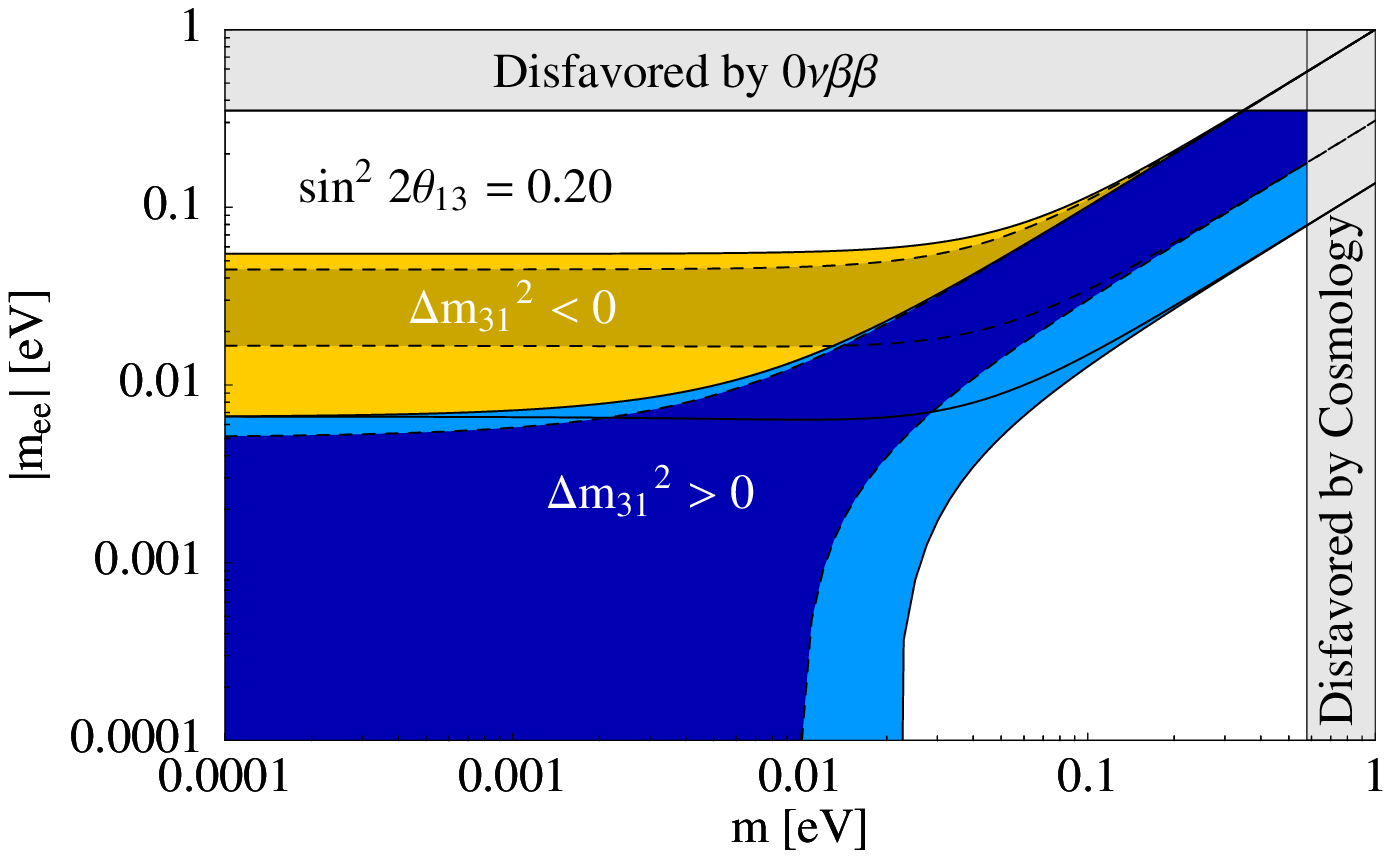,width=8cm,height=6cm}
\caption{\label{fig:usual_boring_plot}The effective mass (in eV) 
for the normal and inverted ordering as a function of the smallest 
neutrino mass (in eV) for different values of $\sin^2 2 \theta_{13}$. 
The prediction for the 
best-fit values of the oscillation parameters and for 
the $3\sigma$ ranges is given. A typical bound from cosmology 
and the limit on the effective mass from Eq.~(\ref{eq:current}) 
are indicated. }
\end{center}
\end{figure}

An interesting aspect is the minimal or maximal value of the effective mass. 
Therefore it is helpful to consider the respective ranges of the three terms 
$|m_{ee}^{(1,2,3)}|$. Maximal \meff\ is obtained when all three 
$|m_{ee}^{(i)}|$ add up, or, in the geometrical picture of 
Fig.~\ref{fig:vectors}, when all three vectors $m_{ee}^{(1,2,3)}$ point 
in the same direction. To find the minimal value of \meff, one has to identify 
the dominating $|m_{ee}^{(i)}|$. In case of 
$|m_{ee}^{(i)}| > |m_{ee}^{(j)}| + |m_{ee}^{(k)}|$, 
the minimal effective mass $\meff_{\rm min}$ is obtained by subtracting the 
two smaller terms from the dominating one. Simply adding or subtracting all 
three terms is equivalent to trivial values of the Majorana phases of 0 or 
$\pi/2$, which corresponds to the conservation of $CP$ \cite{CPmaj}. 
Hence, both the minimal and maximal \meff\ occur in a $CP$ conserving
situation. Note, however, that the Dirac phase which is measurable 
in oscillation experiments can still be non-zero. We introduce the notation 
$--$ to label the case when the second and third term are subtracted from 
the first one. Analogously, the notation for the other two cases is $-+$ 
and $+-$. In Table~\ref{tab:meff_nh_min} we summarize the three possibilities. 
\begin{table}[tb] \hspace{-.7cm}
  \begin{tabular}[tb]{|c|c|c|} \hline \rule[0.5cm]{0cm}{0cm}
  Scenario & Majorana phases & 
$\left|m_{ee}\right|_{\rm min}$ \\[0.2cm] \hline \hline \rule[0.6cm]{0cm}{0cm}
  $\ba \mbox{\small first term dominates} \\  -- \ea $ 
& $\alpha=\beta=\frac{\pi}{2}$ & $\left| m_1 c_{12}^{2} c_{13}^{2}-\sqrt{m_1^{2}+\dms} s_{12}^{2} c_{13}^{2}-\sqrt{m_1^{2}+\dma} s_{13}^{2} \right|$  \\[0.3cm] \hline \rule[0.6cm]{0cm}{0cm}
  $\ba \mbox{\small second term dominates} \\  -+ \ea $ 
 & $\alpha=\frac{\pi}{2},\ \beta=0$ & $\left| m_1 c_{12}^{2} c_{13}^{2}-\sqrt{m_1^{2}+\dms} s_{12}^{2} c_{13}^{2}+\sqrt{m_1^{2}+\dma} s_{13}^{2} \right|$  \\[0.3cm] \hline\rule[0.6cm]{0cm}{0cm}
  $\ba \mbox{\small third term dominates} \\  +- \ea $ 
& $\alpha=0,\ \beta=\frac{\pi}{2}$ & $\left| m_1 c_{12}^{2} c_{13}^{2}+\sqrt{m_1^{2}+\dms} s_{12}^{2} c_{13}^{2}-\sqrt{m_1^{2}+\dma} s_{13}^{2} \right|$  \\[0.3cm] \hline 
  \end{tabular}
 \caption{\label{tab:meff_nh_min}
Minimal values of $\meff$ for dominance of one of the $|m_{ee}^{(i)}|$.}
\end{table}
%

\begin{figure}[tb]
\begin{center}
\input{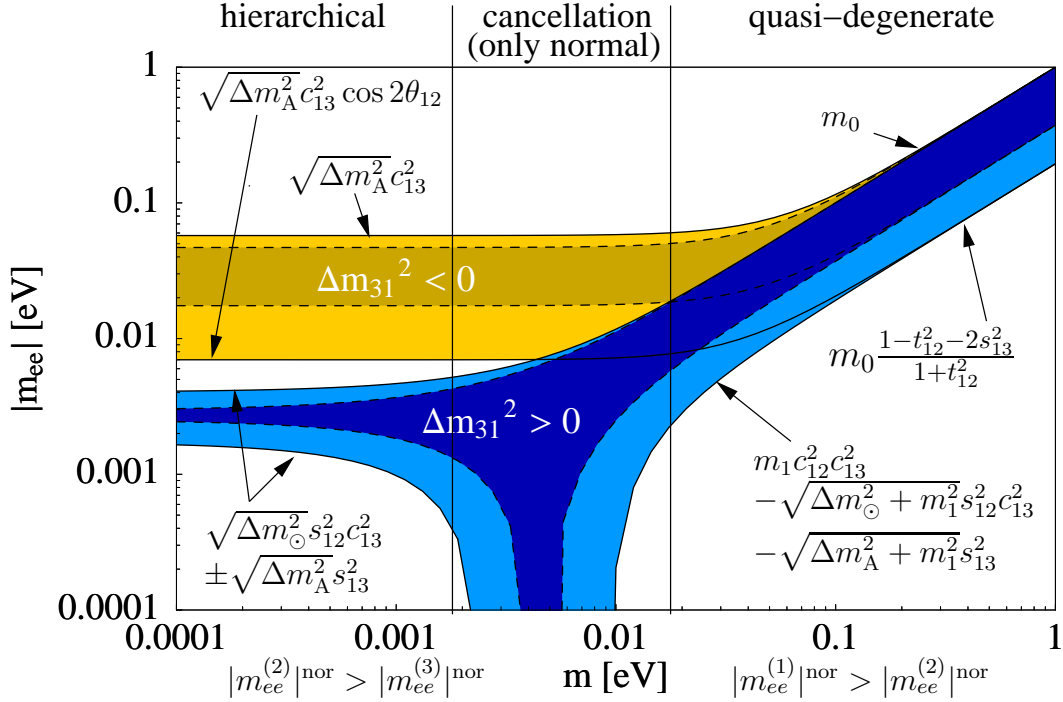}
\caption{\label{fig:lovely_isn't_it?}The main properties 
of the effective mass as function of the smallest neutrino mass. 
We indicated the relevant formulae and the three important regimes: 
hierarchical, cancellation (only possible for normal mass ordering) and 
quasi-degeneracy. 
The value of $\sin^2 2 \theta_{13} = 0.02$ has been chosen, we defined 
$t_{12}^2=\tan^2 \theta_{12}$ and $m_0$ is the common mass scale 
(measurable in KATRIN or by cosmology via $\Sigma/3$) 
for quasi-degenerate neutrino masses 
$m_1 \simeq m_2 \simeq m_3 \equiv m_0$.}
\end{center}
\end{figure}

Using the best-fit and 3$\sigma$ oscillation parameters from
Ref.~\cite{Schwetz}, we can now plot the effective mass as 
a function of the smallest neutrino mass. This is shown in 
Fig.~\ref{fig:usual_boring_plot}, where we assumed different representative 
values of $\theta_{13}$, corresponding to $\sin^2 2 \theta_{13}=0$, 0.03, 
0.1 and 0.2. A typical bound on the sum of neutrino masses
$\Sigma \equiv \sum m_i $ of 1.74~eV is also included (hence $m < 0.58$~eV 
for the lightest neutrino mass), obtained by an analysis of SDSS and 
WMAP data \cite{Tegmark}. 
Moreover, we indicated the limit on the effective mass from 
Eq.~(\ref{eq:current}), where the horizontal line corresponds to 
$\zeta=1$, i.e., everything above the line is unlikely. 
Among the oscillation parameters crucial for \obb, the atmospheric 
$\Delta m^2$ will be known with some precision in the medium future 
\cite{Huber:2004ug}. Generating plots like Fig.~\ref{fig:usual_boring_plot} 
with an assumed error on \dma\ of 10 \% will reveal that the maximal 
effective mass for the inverted ordering is slightly smaller and that 
the minimal effective mass for the normal ordering is slightly larger. 
For our purposes, this will not change the outcome of our conclusions. 

Several features of the figures are immediately identified:
\begin{itemize}
\item[1.)] the effective mass for the normal mass 
ordering can become very small 
or even vanish for certain small values of $m_1$. 
The range of such values of $m_1$ (``the chimney'') 
becomes larger with increasing $\sin^2 2 \theta_{13}$; 
\item[2.)] in case of a normal ordering and a small value 
of $m_1$, the minimal 
value of the effective mass decreases with increasing 
$\sin^2 2 \theta_{13}$; 
\item[3.)]  for small neutrino mass values there is a gap between 
the effective mass in case of a normal and inverted ordering. 
The size of this gap shrinks with increasing $\sin^2 2 \theta_{13}$.  
\end{itemize}
We conclude that there is some interesting interplay between the value of 
$\theta_{13}$ and the effective mass as measurable in \obb. 
In the following, we shall perform a detailed analysis of the effective 
mass for both mass orderings in order to analytically understand 
in particular the features 1.)\ and 2.)\ from above. Then we focus on 
issue 3.)\ and analyze the gap between the minimal value of \meff{} for 
the inverted ordering and the maximal value of \meff{} for the normal  
ordering. In Figure~\ref{fig:lovely_isn't_it?} we show the outcome 
of the coming analysis, taking a typical value of 
$\sin^2 2 \theta_{13} = 0.02$. We indicate the 
relevant regimes and explicitely include the formulae which 
describe the minimal and maximal values of \meff\ in certain ranges.

\section{\label{sec:NH}The Effective Mass for the Normal Mass Ordering}

Let us begin with the normal mass ordering. The effective mass is 
the absolute value of 
\be
m_{ee}^{\rm nor} = m_1 \, c_{12}^{2} \, 
c_{13}^{2} + \sqrt{m_1^{2} + \dms} \, s_{12}^{2} \, c_{13}^{2}\, 
e^{2i\alpha} + \sqrt{m_1^{2} + \dma} \, s_{13}^{2} \, e^{2i\beta} ~.
\label{eq:mee_norm}
\ee
The maximum of the effective mass is obtained when 
the Majorana phases are given by $\alpha=\beta=0$. 
The effective mass is then directly given by the 
real $m_{ee}$:  
\be
\meff^{\rm nor}_{\rm max} = 
m_1 \,  c_{12}^{2} \, 
c_{13}^{2} + \sqrt{m_1^{2} + \dms} \,  s_{12}^{2} \, c_{13}^{2} + 
\sqrt{m_1^{2} + \dma} \,  s_{13}^{2} ~.
\label{eq:meeNHmax}
\ee
Obviously, the largest value of the effective mass is obtained 
when all involved parameters, \dms, \dma, $\theta_{13}$ and $s_{12}^{2}$ 
take their maximally allowed values. For the best-fit, 1 and 3$\sigma$ 
values of the oscillation parameters, the predictions are 
$\meff^{\rm nor}_{\rm max} = 0.10, (0.10, 0.10) \eV$ when $m_1 = 0.1$ eV, 
$\meff^{\rm nor}_{\rm max} = 0.011, (0.012, 0.014)\eV$ when $m_1 = 0.01$ eV, 
and $\meff^{\rm nor}_{\rm max} = 0.0066, (0.0073, 0.0096)\eV$ for 
$m_1 = 0.005$ eV. 

On the other hand, an analytic expression for the minimal value of the 
effective mass is not found easily in every case. Only for very small and rather 
large values of the smallest neutrino mass one can always identify the 
dominating $|m_{ee}^{(i)}|$. A more complicated situation occurs for 
values of \meff{} below roughly $10^{-3}$~eV, i.e., when the effective 
mass is practically zero. 
This interesting region of the plots in Fig.~\ref{fig:usual_boring_plot} 
will be dealt with in detail in Section~\ref{sec:meff_zero}. 
In this case typically two or all three $|m_{ee}^{(i)}|$ are of very 
similar magnitude and small offsets 
in the oscillation parameters or $m_1$ can change the relative 
ordering of the $|m_{ee}^{(i)}|$. 
Some examples for the ranges of the $|m_{ee}^{(i)}|$ are given in 
Table~\ref{tab:meei1ranges} and~\ref{tab:meei3ranges}, inserting the 
1 and 3$\sigma$ oscillation parameters. 
If one of the three $|m_{ee}^{(i)}|$ dominates, we indicated this by 
writing its value in bold face. 
\begin{table}[t]
 \begin{center}
  \begin{tabular}[t]{|c|c|c|c|c|} \hline \rule[0.4cm]{0cm}{0cm}
  $m_1$ [eV] & $\sin^{2} 2\theta_{13}$ & $|m_{ee}^{(1)}|$ [eV] & $|m_{ee}^{(2)}|$ [eV] & $|m_{ee}^{(3)}|$ [eV] \\ \hline \hline \rule[0.4cm]{0cm}{0cm}
  0.1   & 0    & {\bf 0.067--0.072}   & 0.028--0.033    & 0.0000                    \\ \hline \rule[0.4cm]{0cm}{0cm}
        & 0.05 & {\bf 0.066--0.071}   & 0.028--0.033    & 0.0014                    \\ \hline \rule[0.4cm]{0cm}{0cm}
        & 0.2  & {\bf 0.063--0.068}   & 0.027--0.031    & 0.0058--0.0059            \\ \hline \hline \rule[0.4cm]{0cm}{0cm}
  0.01  & 0    & {\bf 0.0067--0.0072} & 0.0037--0.0045  & 0.0000                    \\ \hline \rule[0.4cm]{0cm}{0cm}
        & 0.05 & {\bf 0.0066--0.0071} & 0.0037--0.0044  & (5.7--6.5)$\cdot 10^{-4}$ \\ \hline \rule[0.4cm]{0cm}{0cm}
        & 0.2  & {\bf 0.0063--0.0068} & 0.0035--0.0042  & 0.0024--0.0027            \\ \hline \hline \rule[0.4cm]{0cm}{0cm}
  0.001 & 0    & (6.7--7.2)$\cdot 10^{-4}$ & {\bf 0.0025--0.0030} & 0.000                     \\ \hline \rule[0.4cm]{0cm}{0cm}
        & 0.05 & (6.6--7.1)$\cdot 10^{-4}$ & {\bf 0.0024--0.0030} & (5.6--6.4)$\cdot 10^{-4}$ \\ \hline \rule[0.4cm]{0cm}{0cm}
        & 0.2  & (6.3--6.8)$\cdot 10^{-4}$ & 0.0023--0.0028 & 0.0023--0.0027            \\ \hline \hline \rule[0.4cm]{0cm}{0cm}
 0.0001 & 0    & (6.7--7.2)$\cdot 10^{-5}$ & {\bf 0.0024--0.0030} & 0.0000                    \\ \hline \rule[0.4cm]{0cm}{0cm}
        & 0.05 & (6.6--7.1)$\cdot 10^{-5}$ & {\bf 0.0024--0.0030} & (5.6--6.4)$\cdot 10^{-4}$ \\ \hline \rule[0.4cm]{0cm}{0cm}
        & 0.2  & (6.3--6.8)$\cdot 10^{-5}$ & 0.0023--0.0028 & 0.0023--0.0027            \\ \hline
  \end{tabular}
\caption{\label{tab:meei1ranges}1$\sigma$ ranges of $|m_{ee}^{(i)}|$ for 
different values of $m_1$ and $\theta_{13}$. Bold faced terms indicate 
dominance of the respective term over the whole parameter range.} 
 \end{center}
\end{table}
\begin{table}[t]
 \begin{center}
  \begin{tabular}[t]{|c|c|c|c|c|} \hline \rule[0.4cm]{0cm}{0cm}
  $m_1$ [eV] & $\sin^{2} 2\theta_{13}$ & $|m_{ee}^{(1)}|$ [eV] & $|m_{ee}^{(2)}|$ [eV] & $|m_{ee}^{(3)}|$ [eV] \\ \hline \hline \rule[0.4cm]{0cm}{0cm}
  0.1   & 0    & {\bf 0.060--0.076}              & 0.024--0.040   & 0.0000                   \\ \hline \rule[0.4cm]{0cm}{0cm}
        & 0.05 & {\bf 0.059--0.075}              & 0.024--0.040   & 0.0014--0.0015           \\ \hline \rule[0.4cm]{0cm}{0cm}
        & 0.2  & {\bf 0.057--0.072}              & 0.023--0.038   & 0.0056--0.0061           \\ \hline \hline \rule[0.4cm]{0cm}{0cm}
  0.01  & 0    & {\bf 0.0060--0.0076}            & 0.0031--0.0055 & 0.0000                   \\ \hline \rule[0.4cm]{0cm}{0cm}
        & 0.05 & {\bf 0.0059--0.0076}            & 0.0031--0.0054 & (4.9--7.4)$\cdot 10^{-4}$\\ \hline \rule[0.4cm]{0cm}{0cm}
        & 0.2  & {\bf 0.0057--0.0072}            & 0.0030--0.0052 & 0.0020--0.0031           \\ \hline \hline \rule[0.4cm]{0cm}{0cm}
  0.001 & 0    & (6.0--7.6)$\cdot 10^{-4}$ & {\bf 0.0020--0.0038} & 0.000                    \\ \hline \rule[0.4cm]{0cm}{0cm}
        & 0.05 & (5.9--7.5)$\cdot 10^{-4}$ & {\bf 0.0020--0.0037} & (4.7-7.3)$\cdot 10^{-4}$ \\ \hline \rule[0.4cm]{0cm}{0cm}
        & 0.2  & (5.7--7.2)$\cdot 10^{-4}$ & 0.0019--0.0036 & 0.0020--0.0030           \\ \hline \hline \rule[0.4cm]{0cm}{0cm}
 0.0001 & 0    & (6.0--7.6)$\cdot 10^{-5}$ & {\bf 0.0020--0.0038} & 0.0000                   \\ \hline \rule[0.4cm]{0cm}{0cm}
        & 0.05 & (5.9--7.5)$\cdot 10^{-5}$ & {\bf 0.0020--0.0037} & (4.7--7.3)$\cdot 10^{-5}$\\ \hline \rule[0.4cm]{0cm}{0cm}
        & 0.2  & (5.7--7.2)$\cdot 10^{-5}$ & 0.0019--0.0036 & 0.0020--0.0030           \\ \hline
  \end{tabular}
\caption{\label{tab:meei3ranges}Same as previous Table for the 
$3\sigma$ ranges of the oscillation parameters.}
 \end{center}
\end{table}
With the $1\sigma$ values used in Table~\ref{tab:meei1ranges}, it turns out 
that for very small values of $m_1 \ls 0.001$ eV and 
$\sin^2 2\theta_{13} \ls 0.1$ the term $|m_{ee}^{(2)}|$ always 
dominates\footnote{If one considers the 
extreme case in which \dms\ and $\theta_{12}$ have their 
1(3)$\sigma$ minimum and \dma\ its 1(3)$\sigma$ maximum value, then 
this is not true for $\sin^{2} 2\theta_{13} >$ 0.175~(0.13)
(to be compared with the upper 3$\sigma$-bound of 0.18).}. 
For larger values of $m_1 \gs 0.01$ eV, the term 
$|m_{ee}^{(1)}|$ dominates, irrespective of $\sin^2 2\theta_{13}$. 
These conclusions are rather unaffected by the use of 1 or 3$\sigma$ 
ranges, as can be seen by comparing Tables~\ref{tab:meei1ranges} 
and \ref{tab:meei3ranges}.

\subsection{\label{sec:NHleft}The strictly hierarchical part: $m_1\rightarrow 0$}
Let us focus next on the case of small $m_1$, which 
corresponds to an extreme normal hierarchy (NH), 
defining the ``hierarchical regime'' in Fig.~\ref{fig:lovely_isn't_it?}. 
For small $m_1$ and $\sin^2 2\theta_{13} \ls 0.1$, dominance of 
$|m_{ee}^{(2)}|$ occurs. The effective mass takes its minimal value when $\alpha = \pi/2$ and 
$\beta=0$ ($-+$, see Table~\ref{tab:meff_nh_min}): 
\be
\meff^{\rm nor}_{\rm min} = 
\sqrt{m_1^{2}+\dms} \, s_{12}^2 \, c_{13}^2 - 
m_1 \, c_{12}^{2} \, c_{13}^{2} - \sqrt{m_1^{2}+\dma} \, s_{13}^{2} ~.
\ee
For the best-fit and 1$\sigma$  
values of the oscillation parameters, the predictions are 
$\meff^{\rm nor}_{\rm min} = 0.0021 (0.0011) \eV$ 
when $m_1 = 0.001$ eV and 
$\meff^{\rm nor}_{\rm min} = 0.0024 (0.0015)\eV$ 
when $m_1 = 0.0005$ eV. In this region, we can neglect 
$m_1^2$ with respect to $\dma$. Neglecting 
also $m_1^2$ with respect to $\dms$, we have 
\be
\meff^{\rm nor}_{\rm min, max} \simeq 
\sqrt{\dms} \,  s_{12}^{2} \, c_{13}^{2} \mp \sqrt{\dma} \, s_{13}^{2} ~.
\label{eq:meeNHleft}
\ee
Therefore, for very small values of $m_1$ we expect a comparably 
small band of values of \meff. With increasing $\theta_{13}$, the width 
of the band increases. 
In case of vanishing $\theta_{13}$, we have 
$\meff^{\rm nor}_{\rm min} \simeq \sqrt{\dms} \, s_{12}^{2}$ and the 
band will collapse to a line when \dms{} and $\sin^2 \theta_{12}$ are fixed 
to their best-fit values; the precise value is 2.8~meV. 
All these features are confirmed by Fig.~\ref{fig:usual_boring_plot}.

For finite values of $s_{13}^{2}$, the quantity 
 $|\sqrt{\dms} \, s_{12}^{2} \, c_{13}^{2}- \sqrt{\dma} \, s_{13}^{2}|$ can 
become zero for $s_{13}^2 \simeq s_{12}^2 \, \sqrt{\dms/\dma} 
\simeq 0.034 \ldots 0.090$, where we have inserted the 3$\sigma$ ranges 
of \dms, \dma{} and $\sin^2 \theta_{12}$. 
This range lies partly in the 3$\sigma$ region of $\theta_{13}$. 
For smaller values of $\theta_{13}$, i.e., 
$s_{13}^2 \ls 0.034$, the term $|m_{ee}^{(2)}|$
dominates over $|m_{ee}^{(3)}|$,  
which means that the medium point 
$\sqrt{\dms} \, s_{12}^{2} \, c_{13}^{2}$ of the band is nearly constant 
under variations of $\theta_{13}$, while the width 
$2 \, \sqrt{\dma} \, s_{13}^{2}$ of the band is directly 
proportional to $\left| U_{e3}\right|^{2}$. 
For rather large values of $\theta_{13}$, i.e., 
$s_{13}^2 \gs 0.034$, $|m_{ee}^{(3)}|$ becomes larger than 
$|m_{ee}^{(2)}|$ and the center of the band is at 
$\sqrt{\dma} \, s_{13}^{2}$ and the width is 
$2\sqrt{\dms}\, s_{12}^{2} \, c_{13}^{2}$. 

\subsection{\label{sec:meff_zero}(Nearly) vanishing effective mass}
In the flavor basis, a very small or even vanishing effective 
mass corresponds to a texture zero of the neutrino mass matrix, 
from the theoretical and model building perspective surely 
a highly interesting hint towards the underlying symmetry. 
Fig.~\ref{fig:usual_boring_plot} shows that for not too large values 
of $\sin^2 2 \theta_{13} \ls 0.1$ there is a ``chimney'' of very small 
values of \meff, defining the ``cancellation regime'' in Fig.\ 
\ref{fig:lovely_isn't_it?}. 
Extremely small values of the effective mass are known to have 
interesting phenomenological consequences \cite{small_mee,SC}. 
In the geometrical interpretation of the effective mass, this means that 
the three vectors $m_{ee}^{(1,2,3)}$ can collapse to a triangle. 
In case no single term $|m_{ee}^{(1,2,3)}|$ vanishes (i.e., 
for $m_1 \neq 0$ and $|U_{e3}| \neq 0$) we can apply 
simple geometry (see Fig.~\ref{fig:vectors}) and obtain for $\alpha$
\bea 
\cos 2\alpha = 
\frac{\D |m_{ee}^{(1)}|^{2}+|m_{ee}^{(2)}|^{2}-|m_{ee}^{(3)}|^{2}}
{\D 2 |m_{ee}^{(1)}| \,  |m_{ee}^{(2)}|}= \\[0.5cm] 
\frac{\D m_1^{2} \, \left( c_{13}^{4} \left(s_{12}^{4} + c_{12}^{4} 
\right) - s_{13}^{4} \right) + \dms \, s_{12}^{4} \, c_{13}^{4} - 
\dma \, s_{13}^{4}}
{\D 2 m_1 \, \sqrt{m_1^{2} + \dms} \, s_{12}^{2} \, c_{12}^{2} \, c_{13}^{4}}~,
\eea 
and for $\beta$ 
\bea 
\cos 2\beta = 
\frac{\D |m_{ee}^{(3)}|^{2} + |m_{ee}^{(2)}|^{2} - |m_{ee}^{(1)}|^{2}}
{\D 2 |m_{ee}^{(2)}| \,  |m_{ee}^{(3)}| } \\[0.6cm]
=\frac{\D m_1^{2} \left( s_{13}^{4} \, 
\left( s_{12}^{4} - c_{12}^{4} \right) - s_{13}^{4} \right) 
+ \dms \, s_{12}^{4} \, c_{13}^{4}  + \dma \, s_{13}^{4} }
{\D 2 \sqrt{m_1^{2} + \dms} \, \sqrt{m_1^{2} + \dma} \, 
s_{12}^{2} \, s_{13}^{2} \, c_{13}^{2} }~.
\eea
As interesting, however, is the value of the smallest 
neutrino mass for which the effective mass (nearly) vanishes. 
Let us discuss some special cases: 
\begin{itemize}

\item If $\theta_{13}=0$, then \meff{} vanishes when the 
remaining two terms $m_{ee}^{(1,2)}$ exactly cancel each other 
($\alpha=\pi/2$). For the smallest mass follows: 
\be
m_1 = \tan^{2} \theta_{12} \,  \sqrt{\frac{\dms}{1-\tan^4 \theta_{12}}}
= \sin^{2} \theta_{12} \, \sqrt{\frac{\dms}{\cos 2 \theta_{12}}}
~,
\label{eq:NHminzero13}
\ee
whose best-fit value is 4.5~meV 
(1$\sigma$: 3.7--5.1~meV, 3$\sigma$: 2.8--8.4~meV). 
The width of the ``chimney'' is governed by the range of the relevant 
oscillation parameters. 
For best-fit values (as for any other fixed set of parameters), 
the ``chimney'' is simply a line that crosses the 
zero-$\left| m_{ee}\right|$-axis. 
Its increase after that point is caused by $m_1$ taking values 
larger than the one given in Eq.~(\ref{eq:NHminzero13}) which make 
the mass matrix element $m_{ee}$ switch sign and become negative; 

\item The case of $m_1=0$ was already mentioned in Section~\ref{sec:NHleft}: 
$m_{ee}^{(2,3)}$ have to cancel, or $\alpha=0$, $\beta = \pi/2$ 
(or $\alpha=\pi/2$ and $\beta=0$) and 
consequently the effective mass vanishes if 
\be
\sin^2 2 \theta_{13} = 4 \, 
\frac{\sin^2 \theta_{12} \, \sqrt{\dms}}{\sqrt{\dma} + 
\sin^2 \theta_{12} \, \sqrt{\dms}} \simeq 4 \, 
\sin^2 \theta_{12} \, \sqrt{\frac{\dms}{\dma}} ~,
\ee
whose best-fit value is 0.24  
(1$\sigma$: 0.19--0.28; 3$\sigma$: 0.14--0.40). This effect occurs 
only at rather large values of $\theta_{13}$, as can also be seen in 
Fig.~\ref{fig:usual_boring_plot}; 
 
\item Now we turn to dominance of $m_{ee}^{(2)}$, which is the case for 
small values of $m_1$ and of $\theta_{13}$ 
(neither large $m_1$ nor large $\theta_{13}$ 
should enhance $m_{ee}^{(3)}$). 
With $\sqrt{m_1^{2}+\dma}\simeq \sqrt{\dma}$, the effective mass is 
\be
\meff^{\rm nor}_{\rm min} \simeq \sqrt{m_1^{2} + \dms} \, 
s_{12}^{2} \, c_{13}^{2} - m_1 \, c_{12}^{2} \, 
c_{13}^{2} - \sqrt{\dma} \, s_{13}^{2}  ~. 
\label{eq:2dominates}
\ee
This can be set to zero, and gives with linearizing in 
$m_1$ and using $s_{13}^{4}\simeq 0$:
\be
m_1 \simeq \frac{\dms \, s_{12}^{4}}{2\sqrt{\dma} \, c_{12}^{2} \, 
\tan^{2} \theta_{13}} ~.
\label{eq:m1_2dominates}
\ee
For $\sin^{2} 2 \theta_{13} = 0.02$ the result is 
0.023~(0.016, 0.009) eV, when the oscillation parameters take 
their best-fit and lower 1(3)$\sigma$ values, respectively.  
For $\sin^{2} 2 \theta_{13} = 0.05$ we get 0.0091 eV 
(lower 1$\sigma$: 0.0079 eV, lower 3$\sigma$: 0.0070 eV), 
whereas for $\sin^{2} 2 \theta_{13} = 0.01$ the result is 
0.047~(0.032, 0.019) eV. 
This case is only valid for very specific sets of parameters. 
Therefore we had to insert the lower 1 and 3$\sigma$ values, 
since otherwise the dominance of $m_{ee}^{(2)}$ would be lost;

\item Consider now the case of dominance of $m_{ee}^{(3)}$. 
This situation arises only for rather large values of $\theta_{13}$. 
For the region of the minimum $m_1 \ls 10^{-2}$ eV holds, so that 
by using $\sqrt{m_1^{2}+\dma} \simeq \sqrt{\dma}$, the effective 
mass becomes
\be
\meff^{\rm nor}_{\rm min} \simeq \sqrt{\dma} \, s_{13}^{2} - 
\sqrt{m_1^{2} + \dms} \, s_{12}^{2} \, c_{13}^{2} - m_1 \, c_{12}^{2} \, 
c_{13}^{2}~. 
\label{eq:3dominates}
\ee
Setting this equation to zero, and solving with linearization in $m_1$: 
\be
m_1 \simeq \frac{\dma \, s_{13}^{4} - \dms \, s_{12}^{4} \, c_{13}^{4}}
{2\sqrt{\dma} \, s_{13}^{2} \, c_{13}^{2} \, s_{12}^{2}} = 
\frac{\dma \tan^{2}\theta_{13}-\dms \, s_{12}^{4} \, \cot^{2}\theta_{13}}
{2\sqrt{\dma} \, s_{12}^{2}} ~.
\label{eq:m1_3dominates}
\ee

\end{itemize}

In general, with increasing $\theta_{13}$ the position of the 
minimum shifts towards larger values of $m_1$. Along the same lines, 
for a fixed $m_1$ corresponding to very small \meff, the width of the 
minimum increases with increasing $\theta_{13}$. 
It can also be seen in Fig.~\ref{fig:usual_boring_plot} that 
the smaller the effective mass within this region becomes, 
the smaller the width becomes.
For instance, for $\meff = 10^{-3}$ eV and 
$\sin^2 2 \theta_{13}=0~(0.02,0.2)$, the width is $3.6~(5.0, 12) 
\cdot 10^{-3}$ eV, whereas for $\meff = 10^{-4}$ eV and 
$\sin^2 2 \theta_{13}=0~(0.02,0.2)$, the width is $0.4~(1.7, 10) 
\cdot 10^{-3}$ eV. We used the best-fit oscillation parameters to 
obtain these values. An application of this width is presented in the 
next Subsection.

\begin{table}[tb]
 \begin{center}
  \begin{tabular}[t]{|c|c|c|c|c|} \hline \rule[0.4cm]{0cm}{0cm}
  $\sin^{2}2\theta_{13}$ & $s_{13}^{2}$ & Best-fit & 1$\sigma$ ranges & 3$\sigma$ ranges \\ \hline \rule[0.4cm]{0cm}{0cm}
  0 & 0 & $(5.9-6.5)\cdot 10^{-2}$ eV & $(5.5-6.9)\cdot 10^{-2}$ eV & $(4.7-8.5)\cdot 10^{-2}$ eV\\ \hline \rule[0.4cm]{0cm}{0cm}
  0.03 & 0.008 & $(5.8-6.6)\cdot 10^{-2}$ eV & $(5.4-7.2)\cdot 10^{-2}$ eV & $(4.7-8.9)\cdot 10^{-2}$ eV\\ \hline \rule[0.4cm]{0cm}{0cm}
  0.05 & 0.01 & $(5.8-6.7)\cdot 10^{-2}$ eV & $(5.4-7.3)\cdot 10^{-2}$ eV & $(4.6-9.1)\cdot 10^{-2}$ eV\\ \hline \rule[0.4cm]{0cm}{0cm}
  0.2 & 0.05 & $(5.6-7.6)\cdot 10^{-2}$ eV & $(5.3-8.4)\cdot 10^{-2}$ eV & $(4.5-11.7)\cdot 10^{-2}$ eV\\ \hline
  \end{tabular}
  \caption{\label{tab:cosmo1}Range of $\Sigma$ for $\meff =0.001$ eV.}
 \end{center}
\end{table}
\begin{table}[tb]
 \begin{center}
  \begin{tabular}[t]{|c|c|c|c|c|} \hline \rule[0.4cm]{0cm}{0cm}
  $\sin^{2}2\theta_{13}$ & $s_{13}^{2}$ & Best-fit & 1$\sigma$ ranges & 3$\sigma$ ranges \\ \hline \rule[0.4cm]{0cm}{0cm}
  0 & 0 & $(6.0-6.3)\cdot 10^{-2}$ eV & $(5.6-6.8)\cdot 10^{-2}$ eV & $(4.8-8.2)\cdot 10^{-2}$ eV\\ \hline \rule[0.4cm]{0cm}{0cm}
  0.03 & 0.008 & $(5.9-6.4)\cdot 10^{-2}$ eV & $(5.5-7.0)\cdot 10^{-2}$ eV & $(4.7-8.5)\cdot 10^{-2}$ eV\\ \hline \rule[0.4cm]{0cm}{0cm}
  0.05 & 0.01 & $(5.9-6.5)\cdot 10^{-2}$ eV & $(5.5-7.1)\cdot 10^{-2}$ eV & $(4.7-8.7)\cdot 10^{-2}$ eV\\ \hline \rule[0.4cm]{0cm}{0cm}
  0.2 & 0.05 & $(5.6-7.3)\cdot 10^{-2}$ eV & $(5.3-8.1)\cdot 10^{-2}$ eV & $(4.6-11.1)\cdot 10^{-2}$ eV\\ \hline
  \end{tabular}
  \caption{\label{tab:cosmo2}Range of $\Sigma$ for $\meff =0.0005$ eV.}
 \end{center}
 \begin{center}
  \begin{tabular}[t]{|c|c|c|c|c|} \hline \rule[0.4cm]{0cm}{0cm}
  $\sin^{2}2\theta_{13}$ & $s_{13}^{2}$ & Best-fit & 1$\sigma$ ranges & 3$\sigma$ ranges \\ \hline \rule[0.4cm]{0cm}{0cm}
  0 & 0 & $(6.1-6.2)\cdot 10^{-2}$ eV & $(5.7-6.7)\cdot 10^{-2}$ eV & $(4.9-8.0)\cdot 10^{-2}$ eV\\ \hline \rule[0.4cm]{0cm}{0cm}
  0.03 & 0.008 & $(6.0-6.3)\cdot 10^{-2}$ eV & $(5.6-6.8)\cdot 10^{-2}$ eV & $(4.8-8.2)\cdot 10^{-2}$ eV\\ \hline \rule[0.4cm]{0cm}{0cm}
  0.05 & 0.01 & $(6.0-6.4)\cdot 10^{-2}$ eV & $(5.6-6.9)\cdot 10^{-2}$ eV & $(4.8-8.4)\cdot 10^{-2}$ eV\\ \hline \rule[0.4cm]{0cm}{0cm}
  0.2 & 0.05 & $(5.6-7.1)\cdot 10^{-2}$ eV & $(5.2-7.9)\cdot 10^{-2}$ eV & $(4.6-10.6)\cdot 10^{-2}$ eV\\ \hline
  \end{tabular}
  \caption{\label{tab:cosmo3}Range of $\Sigma$ for $\meff =0.0001$ eV.}
 \end{center}
\end{table}
\begin{figure}[tb]
\begin{center}
\epsfig{file=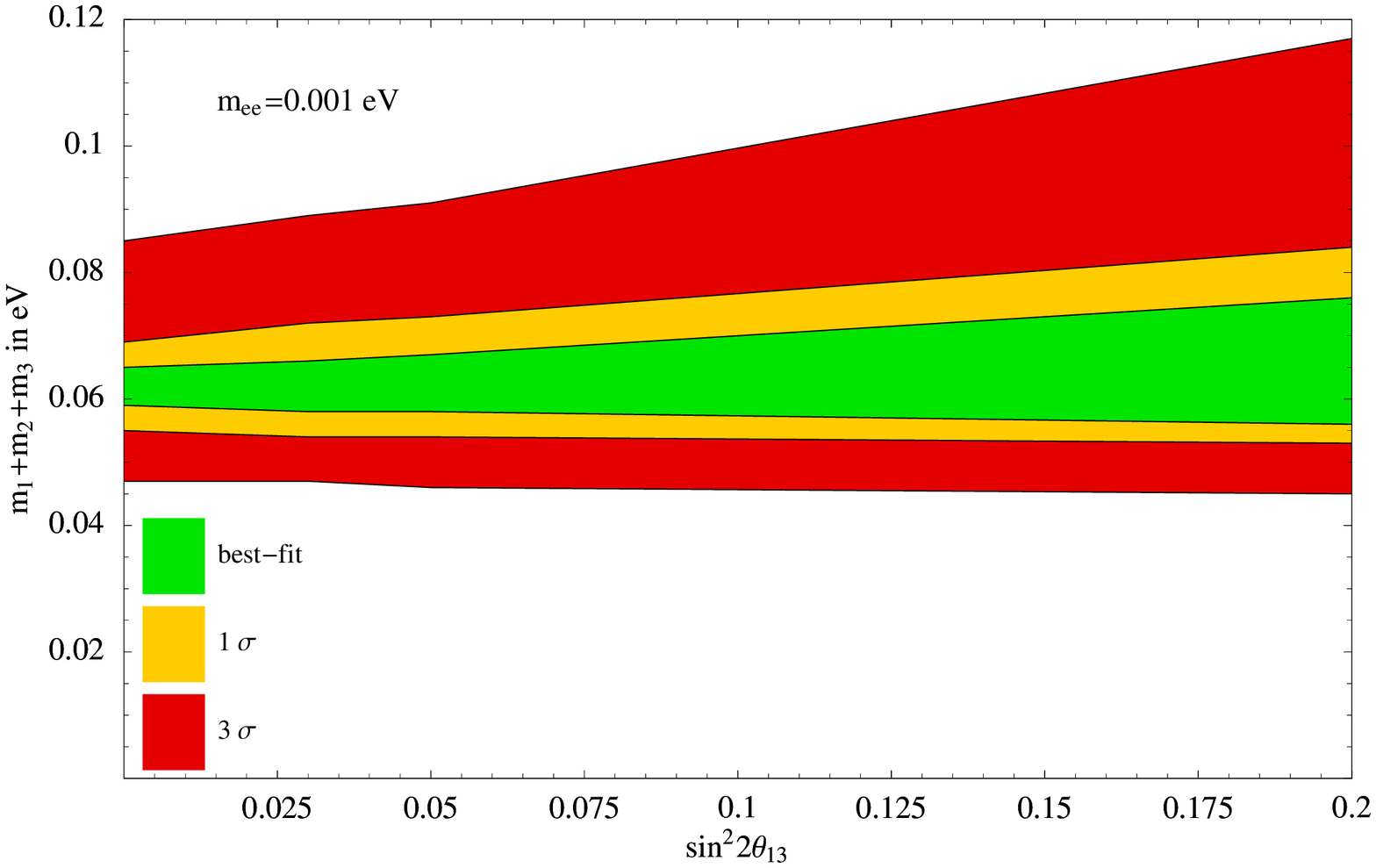,width=8cm,height=6cm}
\epsfig{file=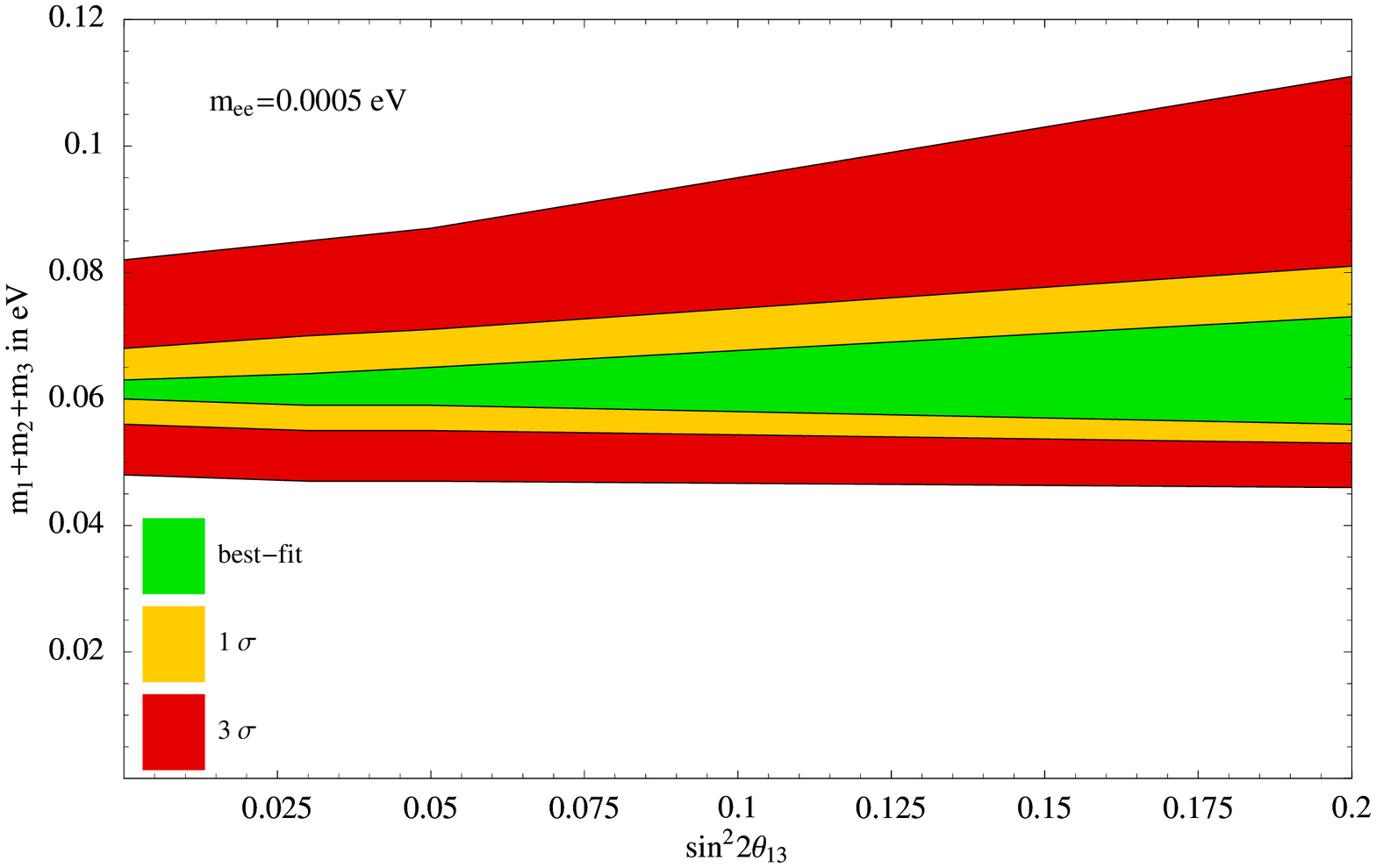,width=8cm,height=6cm}
\epsfig{file=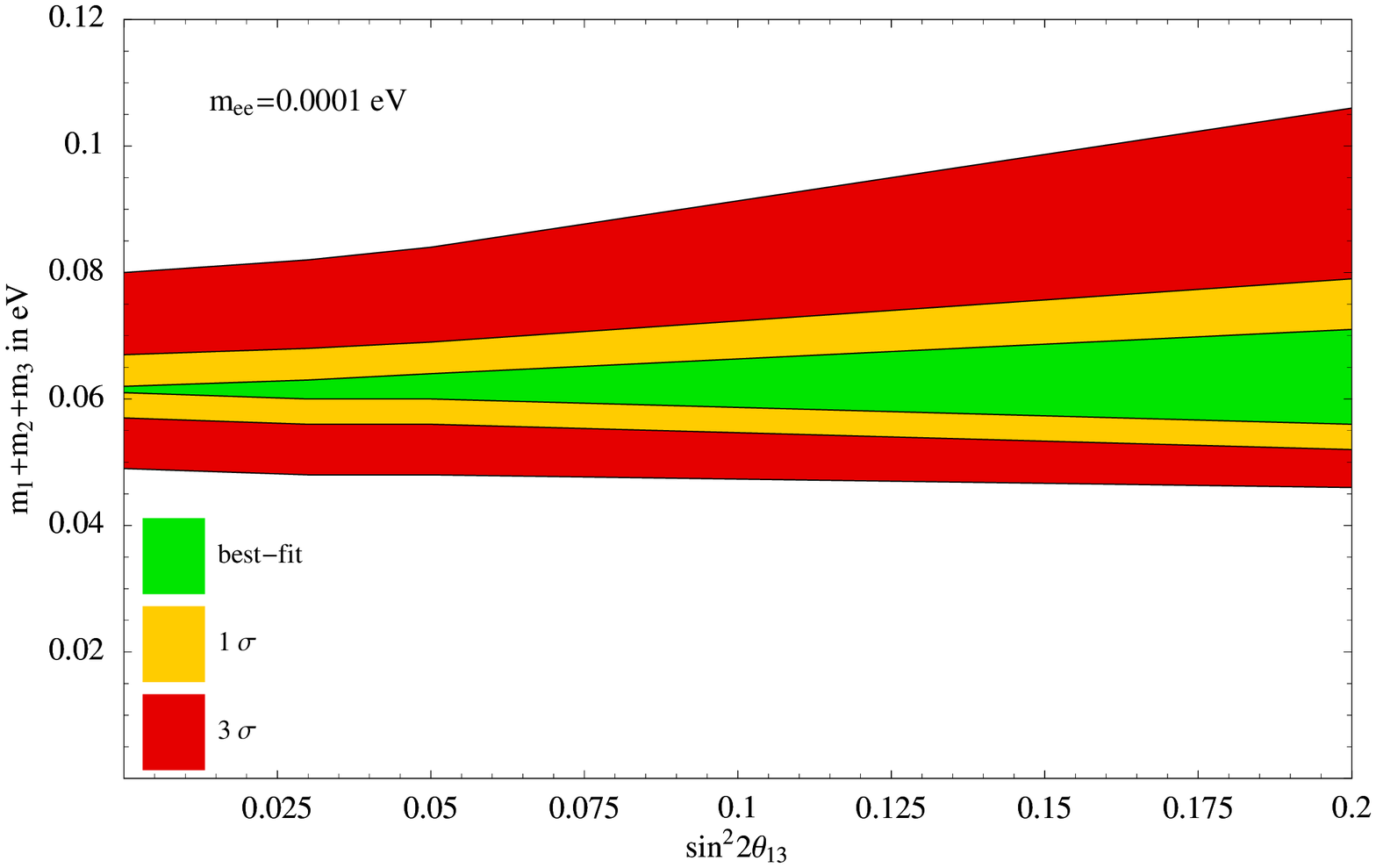,width=8cm,height=6cm}
\epsfig{file=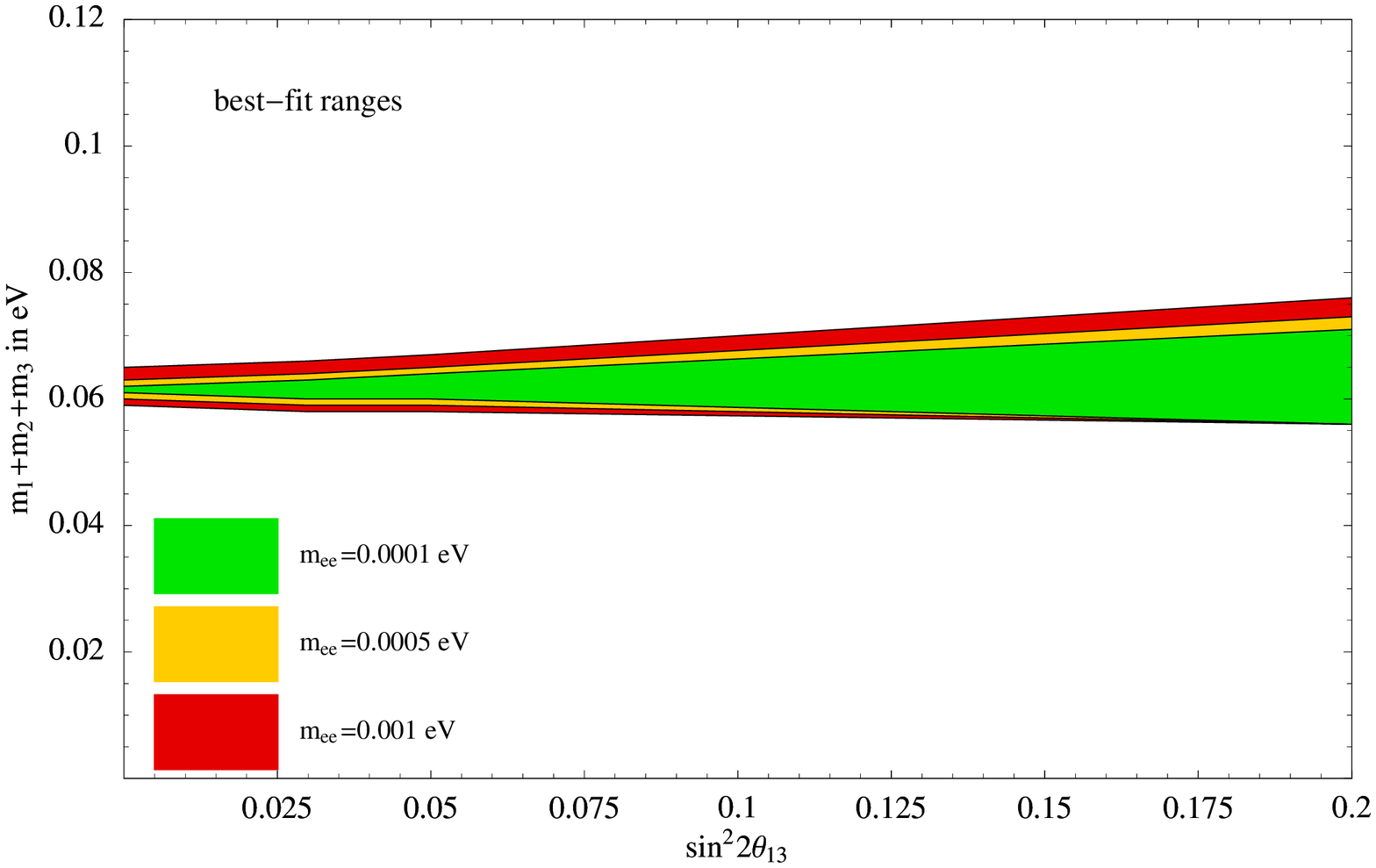,width=8cm,height=6cm}
\caption{\label{fig:cosmo1}The implied values of the sum 
of neutrino masses $\Sigma$ (in eV) for the normal mass ordering 
as a function of $\sin^2 2 \theta_{13}$. Shown are different 
values for \meff~(using the current best-fit, 1 and $3\sigma$ ranges 
of the oscillation parameters).}
\end{center}
\end{figure}

\subsection{\label{sec:cosmo}Interplay with Cosmology for very small \meff}
Let us assume now a very stringent future limit on the effective mass. 
The only interpretation of this hypothetical, but also 
realistic, situation is then that the smallest neutrino mass takes 
values within the ``chimney'' corresponding to 
extremely small values\footnote{Of course, neutrinos could then 
simply be Dirac particles. Let us however not bother about this dreadful 
possibility any more.} of \meff. 
Moreover, the normal mass ordering has to be present, an assertion 
that might at that point of time already have been confirmed by an 
independent oscillation experiment. 
With the indicated values of $m_1$, we can go on and calculate the 
sum of neutrino masses, 
\be
 \Sigma = m_1 + m_2 + m_3 = m_1 + \sqrt{m_1^{2}+\dms} + \sqrt{m_1^{2}+\dma} ~,
\label{eq:cosmologyM}
\ee
because it is this very quantity, which will also witness some improvement 
regarding our knowledge about it \cite{steen}. 
Using the current 3$\sigma$ ranges of \dms\ and \dma, and some 
values of $\theta_{13}$, gives the ranges for $\Sigma$ displayed 
in Tables~\ref{tab:cosmo1} to \ref{tab:cosmo3} and in Fig.~\ref{fig:cosmo1}.

One can read off the Tables and the 
Figure that $\Sigma$ is around 0.1 eV and that its upper limit 
moderately increases with $\theta_{13}$.  
Recall that -- as shown in the previous Subsection -- 
the width of the ``chimney'' grows with 
$\theta_{13}$. The major effect of broadening of the ranges of $\Sigma$ 
comes from the variation of the oscillation parameter ranges and, 
as can be seen from the plot with their values fixed to the 
best-fit values, not from the exact upper limit on $\meff$. 
Hence, having a limit of 0.001 eV on the effective mass is enough to 
reach the implied values of $\Sigma$ around 0.1 eV.

The current limit on the sum of neutrino masses lies between 
0.42 eV \cite{seljak} and 1.8 eV \cite{Tegmark}, depending on the data sets 
and priors used in the analysis. Future improvement of one 
order of magnitude is discussed in the literature \cite{steen}. 
Consider now a limit on the effective mass of 0.001 eV. 
Then, the implied $1\sigma$ range of $\Sigma$ 
(with such a small limit on \meff, the errors on the oscillation 
parameters are expected to be small, too) 
is between roughly 0.055 and 0.08 eV. The conservative limit 
on $\Sigma < 1.8$ eV has to be improved merely by a factor of 
20 to 40 to fully probe this region. 
We note finally that a determination of the effective mass above 0.001 
eV will lead to testable consequences for cosmology anyway (see
e.g.~\cite{others}). Here we wish to stress that even a negative 
search for \meff\  has some testable impact on cosmology.

\subsection{\label{sec:NHdegereration}Transition to the 
quasi-degenerate region}
For larger neutrinos masses corresponding to $m_1 \gs 0.03$ eV, the neutrino masses 
perform a transition to the ``quasi-degenerate regime'' in 
Fig.~\ref{fig:lovely_isn't_it?}, i.e., 
corrections to $m_3 = m_2 = m_1$ are sub-leading. 
The mass matrix element is given by 
\be
m_{ee}^{\rm nor} \simeq m_1 \, 
\left( 
c_{12}^{2} \, c_{13}^{2} + s_{12}^{2} \, c_{13}^{2} \, e^{2i \alpha} 
+ s_{13}^{2} \, e^{2i \beta}
\right) ~.
\ee
The effective mass scales with $m_1$, which in this regime is also 
the neutrino mass measured in kinematical searches such as KATRIN 
(in cosmological searches, it would also appear at $m_1 \simeq \Sigma /3$). 
In fact, the maximal value of \meff\ is nothing but $m_1$. 
It holds now 
$|m_{ee}^{(3)}| \ll |m_{ee}^{(2)}| < |m_{ee}^{(1)}|$ and 
therefore the minimal value of \meff{} is 
given by subtracting the second and 
third term from the first one, or $\alpha = \beta = \pi/2$ 
($--$, see Table~\ref{tab:meff_nh_min}):
\bea
\meff^{\rm nor}_{\rm min} = 
 m_1 \, c_{12}^{2} \, c_{13}^{2} - 
\sqrt{m_1^{2}+\dms} \, s_{12}^{2} \, c_{13}^{2} - 
\sqrt{m_1^{2}+\dma} \, s_{13}^{2} \\[0.2cm]
\simeq m_1 \, \left( 
|U_{e1}|^2 - |U_{e2}|^2 - |U_{e3}|^2 
\right) = m_1 \, \frac{\D 1 - \tan^2 \theta_{12} - 2 \, 
\sin^2 \theta_{13} }
{\D 1 + \tan^2 \theta_{12}} \equiv m_1 \, f(\theta_{12}, \theta_{13})~.
\label{eq:meeQDapp}
\eea
The function $f(\theta_{12}, \theta_{13})$ \cite{SC} introduced 
in this equation has a best-fit value of $0.38$ and a 1(3)$\sigma$ range of 
$0.32$--$0.44$ ($0.15$--$0.52$). 
The quantity $m_1(1 - f(\theta_{12}, \theta_{13}) )$ 
defines the width of the band in the quasi-degenerate regime 
in Fig.~\ref{fig:lovely_isn't_it?}.

\section{\label{sec:IH}The Effective Mass for the Inverted Mass Ordering}

For the inverted mass ordering, the smallest neutrino mass is denoted $m_3$ 
and the mass matrix element is given by 
\be
m_{ee}^{\rm inv} = \sqrt{m_3^{2}+\dma} \,  c_{12}^{2} \, 
c_{13}^{2} + \sqrt{m_3^{2} + \dms + \dma} \,  s_{12}^{2} \, c_{13}^{2} \, 
e^{2i\alpha} + m_3 \,  s_{13}^{2} \, e^{2i\beta} ~.
\label{eq:meeIH}
\ee
The maximal effective mass is -- as for the normal mass ordering -- 
obtained by adding the three terms: 
\be
\left|m_{ee} \right|^{\rm inv}_{\rm max} = 
\sqrt{m_3^{2}+\dma} \, c_{12}^{2} \, c_{13}^{2} 
+ \sqrt{m_3^{2} + \dms + \dma} \, s_{12}^{2} \, c_{13}^{2} + m_3 \, 
s_{13}^{2} ~.
\label{eq:meeIHmax}
\ee
Finding the minimal \meff{} is rather easy. 
With  $\dma \gg \dms$ one gets for all $m_3$ 
$$\frac{|m_{ee}^{(2)}|}{|m_{ee}^{(1)}|} \simeq \tan^{2} \theta_{12}~~
\mbox{ and }\ \ \frac{|m_{ee}^{(3)}|}{|m_{ee}^{(2)}|} = 
\frac{m_3}{\sqrt{m_3^2 + \dma}} \, \frac{s_{13}^2 }{c_{12}^2 \, c_{13}^2}~,
$$ 
which shows that $|m_{ee}^{(2)}/m_{ee}^{(1)}|$ is always smaller   
and $|m_{ee}^{(3)}/m_{ee}^{(2)}|$ always much smaller than one. 
Hence, for all values of $m_3$ we have  $|m_{ee}^{(3)}| \ll |m_{ee}^{(2)}| < 
|m_{ee}^{(1)}|$ and the minimal value of \meff{} is obtained by 
subtracting $|m_{ee}^{(3)}|$ and $|m_{ee}^{(2)}|$ 
from $|m_{ee}^{(1)}|$, i.e., by choosing 
$\alpha=\beta=\frac{\pi}{2}$ ($--$, see Table~\ref{tab:meff_nh_min}):  
\be
\meff^{\rm inv}_{\rm min} 
=\sqrt{m_3^{2}+\dma}\, c_{12}^{2} \, c_{13}^{2} - 
\sqrt{m_3^{2} + \dms + \dma}\, s_{12}^{2} \, c_{13}^{2} - m_3\, s_{13}^{2} ~.
\label{eq:meeIHmin}
\ee
The equations (\ref{eq:meeIHmax}) and (\ref{eq:meeIHmin})
define the upper and the lower line of the band in Fig.\ 
\ref{fig:usual_boring_plot}.

The largest possible \meff{} is obtained for the largest values of 
\dma\, \dms\ and $s_{12}^{2}$ as well as for the smallest value of $s_{13}^{2}$. 
In fact, the dependence of  $\meff^{\rm inv}_{\rm max}$  
on $s_{12}^{2}$ is small, since this parameter enters only via 
$c_{13}^{2} \, \left(\sqrt{m_3^{2}+\dma+\dms}-\sqrt{m_3^{2}+\dma}\right) \, 
s_{12}^{2}$, which is only of order $\dms/\sqrt{\dma + m_3^2}$.   
The smallest value of $\meff^{\rm inv}_{\rm max}$ 
is reached for the largest \dms, $s_{13}^{2}$ and $s_{12}^{2}$ 
as well as the smallest \dma.

\subsection{\label{sec:IHleft}The strictly hierarchical part: 
$m_3\rightarrow 0$}

One important case is that of a vanishing lightest neutrino mass, i.e., 
$m_3\rightarrow 0$, the hierarchical regime in Fig.\ 
\ref{fig:lovely_isn't_it?}. 
In this case \cite{others,SC,SPP},
\bea
m_{ee}^{\rm inv} \simeq \sqrt{\dma}\,c_{13}^{2}  
\left( c_{12}^{2} + s_{12}^{2}  \, e^{2i\alpha} \right) \\[0.2cm]
\mbox{ and } 
\meff^{\rm inv}_{\rm max} \equiv \sqrt{\dma}\,c_{13}^{2} \ge 
\meff^{\rm inv} \ge \sqrt{\dma}\,c_{13}^{2} \cos 2 \theta_{12} 
\equiv \meff^{\rm inv}_{\rm min}~.
\label{eq:meeIHleft}
\eea
From this formula one can see that even for vanishing $s_{13}^{2}$  
the band for small neutrino masses 
has -- in contrast to the normal mass ordering -- a certain width, 
given by the allowed range or value of 
$2 \, \sqrt{\dma} \, \sin^2 \theta_{12}$. 
For best-fit values (1, 3 $\sigma$ ranges), 
the width is 0.03 eV (between 0.025 eV and 0.034 eV, 0.018 eV and 0.046 eV, 
respectively).  
The dependence on $s_{13}^{2}$ is rather small for the inverted mass ordering, 
and the effective mass contains information mainly on $\dms$, 
$\sin^2 \theta_{12}$ and, in principle, on one of the Majorana phases.

\subsection{\label{sec:IHdegereration}Transition to the 
quasi-degenerate region}

The transition to the quasi-degenerate regime takes place when 
$m_3 \gs 0.03$ eV. 
If the smallest mass assumes such values, the normal and inverted mass 
ordering generate identical predictions for the effective mass. 
The results in this case are therefore identical to the ones for the normal mass 
ordering treated above in Section~\ref{sec:NHdegereration} and can 
be obtained by replacing $m_1$ with $m_3$ in the formulae.

\section{\label{sec:combined}Normal vs.\ Inverted Mass Ordering}

Having discussed the normal and inverted mass ordering in some detail, 
we can turn now to a very important aspect of \obb, namely the 
possible distinction of the mass orderings \cite{NHvsIH,SC,SPP}. 
As we have argued in Section~\ref{sec:mee}, 
the gap between the inverted and normal mass ordering for small masses, 
i.e., for IH and NH, enjoys some dependence on the value of 
$\theta_{13}$. 
By glancing at Fig.~\ref{fig:usual_boring_plot} 
or \ref{fig:lovely_isn't_it?}, we see 
that the gap between NH and IH depends also on the 
precision of the oscillation parameters. For the $3\sigma$ values there 
is a gap for neutrino masses below a few $10^{-3}$ eV, whereas 
the best-fit values allow a distinction for neutrino masses below 
roughly $10^{-2}$ eV. Of course, it is the value of 
$\theta_{12}$ which plays the main role here \cite{SC}. 
Another point of concern is the uncertainty generated by different 
calculations of the nuclear matrix elements, which has to be taken into 
account now.

To do that, we call the nuclear matrix element uncertainty 
$\zeta$. We have to 
calculate the difference between the minimal effective mass 
for the inverted ordering and the maximal effective mass for the 
normal ordering multiplied with the uncertainty factor $\zeta$: 
\be \label{eq:Delta}
\Delta \meff \equiv \meff^{\rm inv}_{\rm min} - \zeta \, 
\meff^{\rm nor}_{\rm max} ~.
\ee 
The maximal value of 
$\meff^{\rm nor}$ is given in Eq.~(\ref{eq:meeNHmax}), and the minimal 
value of $\meff^{\rm inv}$ in Eq.~(\ref{eq:meeIHmin}). 
Denoting the smallest neutrino mass with $m_{\rm sm}$, we have 
in general 
\be
\Delta \meff =
\meff^{\rm inv}_{\rm min} - \zeta \, 
\meff^{\rm nor}_{\rm max} 
=  \left(\sqrt{m_{\rm sm}^{2} + \dma} - \zeta \, m_{\rm sm} \right) 
\, c_{12}^{2} \, c_{13}^{2}-\nonumber \\
\left(\sqrt{m_{\rm sm}^{2} + \dms + \dma} + \zeta \, 
\sqrt{m_{\rm sm}^{2} + \dms} \right)  s_{12}^{2} \, c_{13}^{2} - 
\left(\zeta \, \sqrt{m_{\rm sm}^{2} + \dma} + m_{\rm sm} \right) 
 s_{13}^{2} ~.
\label{eq:gap}
\ee
The indicated value of $\Delta \meff$ represents the maximal experimental 
uncertainty in the determination of \meff\ \cite{SC}. 
For larger uncertainties, 
distinguishing NH from IH becomes impossible.

\begin{figure}[tb]
\begin{center}
\epsfig{file=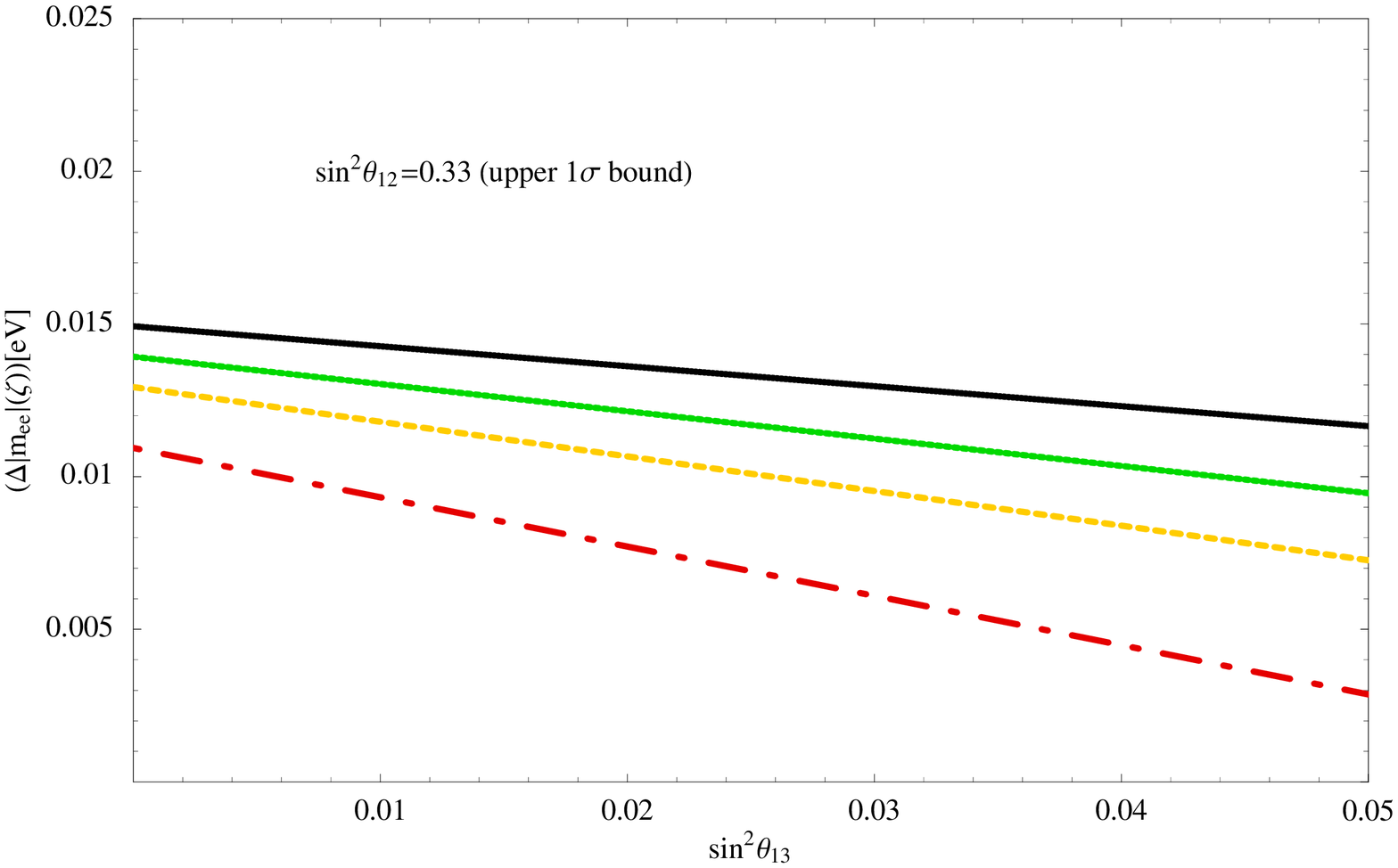,width=8cm,height=6cm}
\epsfig{file=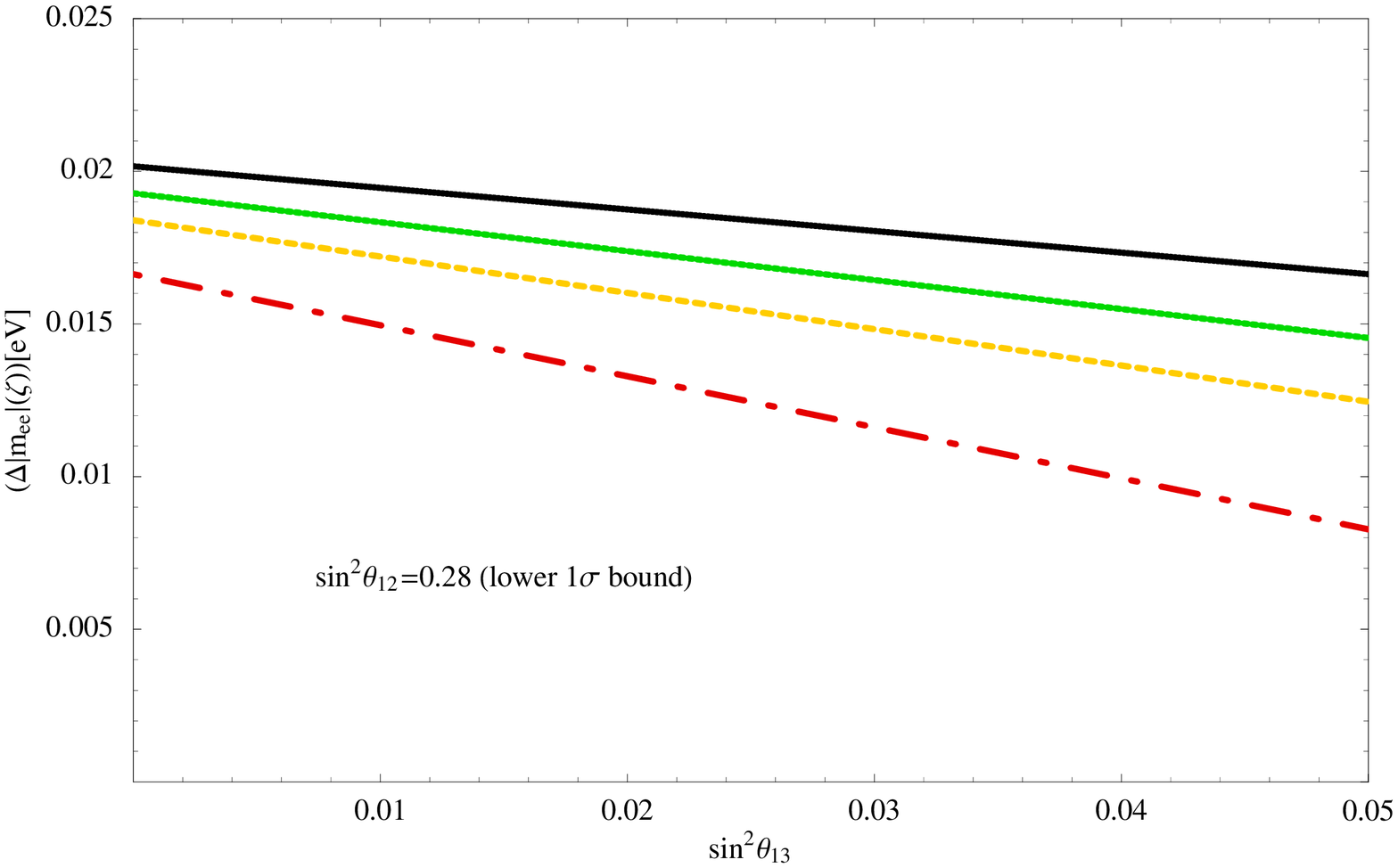,width=8cm,height=6cm}
\epsfig{file=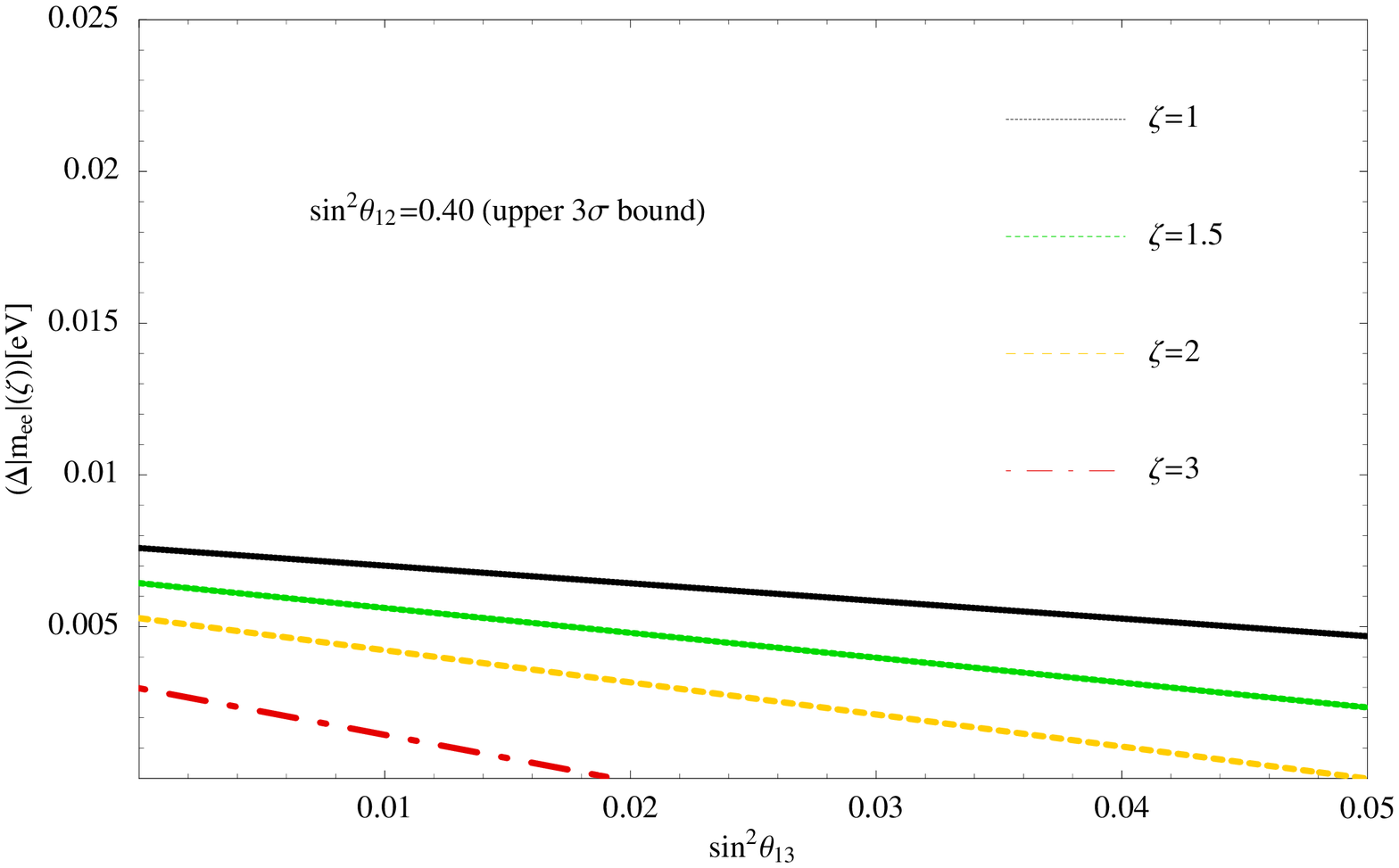,width=8cm,height=6cm}
\epsfig{file=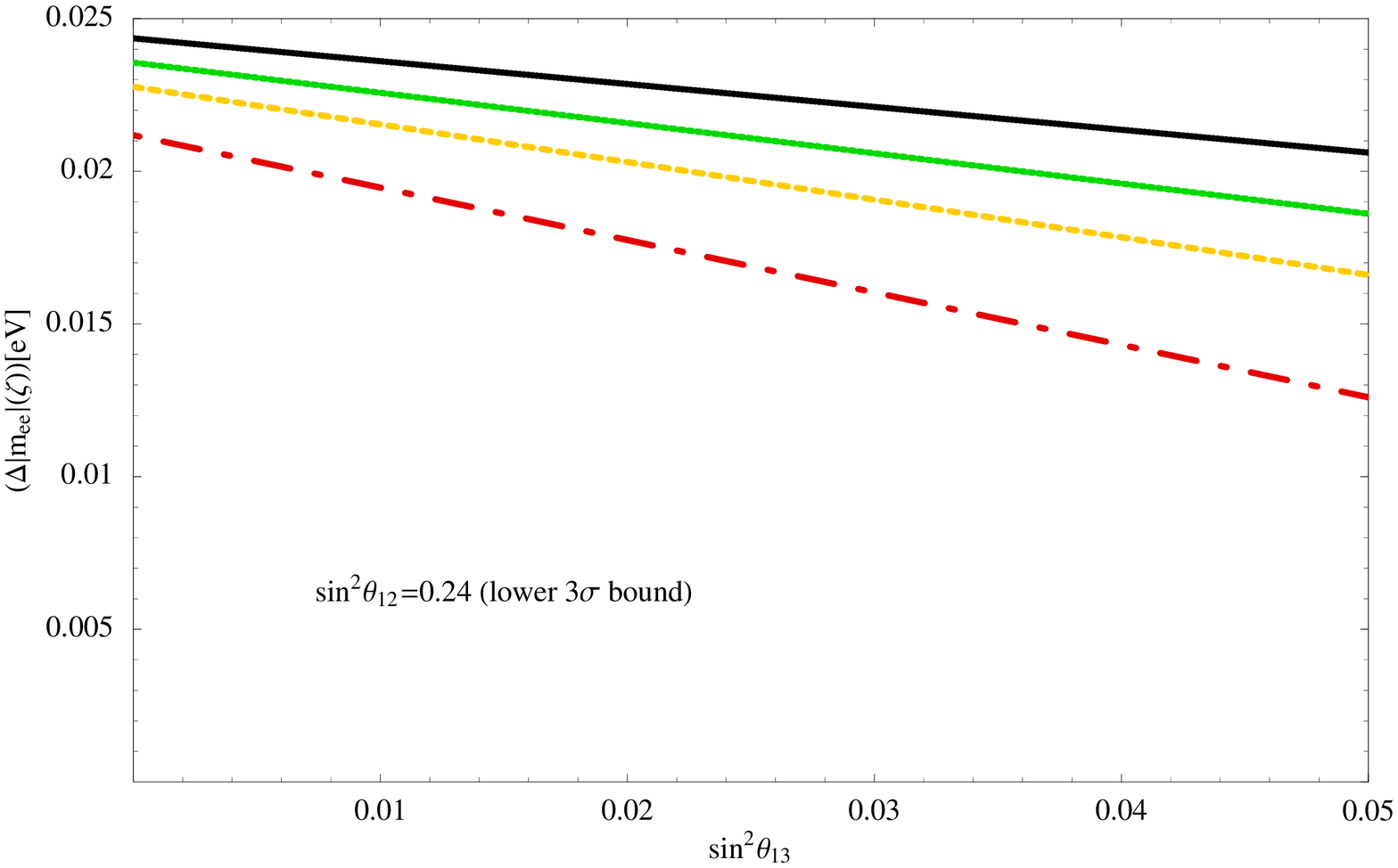,width=8cm,height=6cm}
\caption{\label{fig:gaps}The difference $\Delta \meff$ of $\meff^{\rm inv}_{\rm min}$ and  
$\zeta \, \meff^{\rm nor}_{\rm max} $ as a function of 
$\sin^2 2 \theta_{13}$ for an illustrative value of 
$m_{\rm sm}=0.005$ eV, different sets of oscillation parameters and 
different nuclear matrix element uncertainty factors $\zeta$. }
\end{center}
\end{figure}

The variation of $\Delta \meff$ with $\theta_{13}$ is only slow, 
as a function of $s_{13}^{2}$ or $\sin^2 2 \theta_{13}$ it is basically 
a monotonously decreasing line starting from a value roughly given by 
$\meff^{\rm inv}_{\rm min}$. 
The value of $\zeta$ effectively increases the negative slope of this line.

The order of magnitude of $\Delta \meff$ is generically 
$\sqrt{\dma} \, \cos^2 \theta_{13}$. This can be seen, for instance, 
when we define the small quantities 
\[
R \equiv \frac{\dms}{\dma} ~~\mbox{ and }~~ \eta \equiv 
\frac{m_{\rm sm}}{\sqrt{\dma}}~,
\]
which allow to rewrite Eq.~(\ref{eq:gap}) as 
\bea
\Delta \meff = \sqrt{\dma} \, c_{13}^2 \, 
\left(
c_{12}^2 \, \left( \sqrt{1 - \eta^2} - \eta \, \zeta \right) 
- s_{12}^2 \, 
\left( \sqrt{1 - \eta^2 + R} + \zeta \, \sqrt{R + \eta^2 }\right) 
\right. \\[0.2cm]
\left. - \left( \eta + \sqrt{1 - \eta^2} \, \zeta \right) \, \tan^2 \theta_{13}
\right)~.
\eea
At zeroth order in all small quantities $R$, $\eta$ and $\theta_{13}$, 
we have $\Delta \meff \simeq \sqrt{\dma} \, \cos 2 \theta_{12}$, 
which is nothing but $\meff^{\rm inv}_{\rm min}$.

Using $\dma+\dms \simeq \dma$ and taking the limit 
$m_{\rm sm} \rightarrow 0$, we get from Eq. (\ref{eq:gap})
\be
\Delta \meff(m_{\rm sm} \rightarrow 0) \simeq 
\sqrt{\dma} \left(c_{13}^{2} \, (c_{12}^{2} - s_{12}^{2}) 
- \zeta \, s_{13}^{2} \right) - \zeta \, 
\sqrt{\dms} \, s_{12}^{2} \, c_{13}^{2}~.
\label{eq:gapzerom}
\ee
For no uncertainty, i.e., if $\zeta=1$, and for the 
current best-fit values of the oscillation parameters, 
this function monotonously decreases from 15.0~meV 
for $\theta_{13}=0$ to 12.0~meV for $s_{13}^{2}=0.05$. 
If $\zeta=2$, then it decreases from 12.3~meV 
for $\theta_{13}=0$ to 7.0~meV for $s_{13}^{2}=0.05$. 
As noted in Ref.\ \cite{SC}, the dependence on $\theta_{12}$ 
of $\Delta \meff$ is rather strong. 
We give a few numerical examples, obtained for a vanishing 
smallest neutrino mass: 
if we take $\zeta=1$ and 
$\sin^2 \theta_{12} = 0.24$ (lower 3$\sigma$ value), 
then $\Delta \meff$ decreases from 22.3~meV for 
$\theta_{13}=0$ to 18.8~meV for $s_{13}^{2}=0.05$. 
For the same $\sin^2 \theta_{12}$ and $\zeta=2$, its values are 
14.4~meV ($\theta_{13}=0$) and 5.8~meV ($s_{13}^{2}=0.05$). 
For $\sin^2 \theta_{12} = 0.40$ (upper 3$\sigma$ value) in turn, 
$\Delta \meff$ decreases from 5.8~meV for $\theta_{13}=0$ to 3.2~meV 
for $s_{13}^{2}=0.05$ if $\zeta=1$, while for $\zeta=2$ it 
starts at 2.3~meV and crosses zero for $s_{13}^{2} \simeq 0.024$. 
Values of $\Delta \meff$ equal to or less than zero mean that one 
cannot distinguish the normal from the inverted hierarchy anymore. 

For $\theta_{13}=0$ the variation of the oscillation parameters 
gives a range of $\Delta \meff$ from 12 to 20~meV ($1\sigma$) 
or 4 to 28~meV ($3\sigma$) for $\zeta=1$ and from 9 to 17~meV 
($1\sigma$) or 0 to 26~meV ($3\sigma$) for $\zeta=2$ 
(within the parameter range of the oscillation parameters 
$\Delta \meff$ can become less than zero). 
Fixing the oscillation parameters to their best-fit values 
and varying $\zeta$ from 1 to 5 leads to a range of $\Delta \meff$ 
from 15 to 4~meV. For $\sin^2 2 \theta_{13} = 0.02$~(0.2) 
the range is 14.8 to 2.8~(11.8 to 0)~meV. 

For an illustrative value of $m_{\rm sm}=0.005$ eV and for different 
$\sin^2 \theta_{12}$ and $\zeta$ we show $\Delta \meff$ as a function of 
$\sin^2  \theta_{13}$ in Fig.~\ref{fig:gaps}. We see that if the true 
value of $\sin^2 \theta_{12}$ is not too far away from its current 
best-fit value and if $\zeta \ls 2$, 
then $\Delta \meff$ lies always around 0.01 eV unless 
$\theta_{13}$ is very close to its current upper limit. 
If $\sin^2 \theta_{12}$ is 
on the upper side of its allowed range or $\zeta \gs 2$, then rather 
small values of $\Delta \meff$ are implied. We remark that recent 
investigations seem to indicate that indeed $\zeta \ls 2$ \cite{NME}.\\ 

An interesting point worth stressing is the complementary role played by 
\obb{}  and oscillation experiments in what regards the 
determination of the neutrino mass hierarchy. 
As we discussed here in some detail, the gap $\Delta \meff$ 
between IH and NH decreases for increasing values of $\theta_{13}$. 
For oscillation experiments on the other hand, one typically uses 
matter effects on $\theta_{13}$ to pin down the hierarchy. 
Consequently, in case of zero $\theta_{13}$ these efforts are doomed. 
In principle it will still be possible to determine the hierarchy in 
oscillation experiments, 
but this typically requires a precision measurement of \dma{} on a level 
of \dms{} \cite{LBL_new}, which is quite challenging. 
Hence, the larger $\theta_{13}$, the easier it will be to measure the 
mass ordering, i.e., the sign of $m_3^2 - m_1^2$. 
From this point of view, both types of experiments are complementary. 
Let us however not shut off from view that the identification of the 
sign of $m_3^2 - m_1^2$ via \obb\ depends on the fact that 
the smallest neutrino mass indeed should be small, say, $m_{\rm sm} \ls 0.01$ eV. 
However, most GUT based models predicting neutrino parameters 
predict a normal hierarchy with such light neutrino masses 
(for a summary of possibilities and models, see for instance 
\cite{APSgen,SC}).  
If a model incorporates the inverted mass ordering, then stability 
under radiative corrections demands 
usually the flavor symmetry $L_e - L_\mu - L_\tau$ \cite{STP} 
to play a role, and 
consequently, even after breaking the symmetry, the smallest mass is very light, too.

Moreover, any extraction of information 
from \obb{} has some intrinsic model dependence. The most important one 
is the assertion that neutrinos are Majorana particles, which however 
has more than only solid theoretical foundation. Then again, 
there are several diagrams 
of Physics beyond the Standard Model which in principle can mediate 
\onbb. However, no such New Physics candidate has shown up so far, and the 
indisputable evidence for neutrino oscillations indicates that the 
neutrino-mass-mediated channel of \obb\ is present. Since any 
Feynman diagram leading to \obb\ automatically generates a 
(loop-suppressed) Majorana mass term for the neutrinos \cite{scheval}, 
one would have to explain why massive neutrino are a sub-leading contribution 
to \obb\ but the other New Physics responsible for it 
does not show up elsewhere.

\section{\label{sec:concl}Conclusions}

Future measurements will improve the sensitivity for $\sin^22\theta_{13}$ 
by at least one order of magnitude within the next years. At the same 
time there will be considerable improvements in the determination of 
the absolute neutrino mass scale from neutrino-less double beta decay 
and from cosmology. We discussed in this paper the interplay of these 
improvements. 
Especially, we showed that a measurement or an improved limit 
of $\theta_{13}$ is very important for the separation and for the precise 
form of the normal and inverted hierarchy solution for hierarchical 
neutrino masses. We demonstrated that for todays largest possible values 
of $\theta_{13}$, the normal and inverted hierarchy regions overlap. 
An improvement of $\theta_{13}$ is especially important to be able to 
fully exclude or probe the inverted mass hierarchy with next generation 
of \obb~experiments like GERDA, CUORE or MAJORANA. In addition, we showed 
that in the case of a normal hierarchy, arbitrarily small values of the 
effective neutrino mass are allowed for these largest possible values 
of $\theta_{13}$. 
For intermediate values of the absolute neutrino mass scale we showed 
that the width of the ``chimney'' depends sizably on $\theta_{13}$.
The anticipated improvement by one order of magnitude will make this
``chimney'' rather narrow. Even though the chimney exists for arbitrarily
small values of $\theta_{13}$, its width becomes so narrow that it 
would correspond to rather specifically chosen parameter values. 
If \obb~experiments reach a sensitivity for $|m_{ee}|\ls 10^{-3}$~eV,
and if neutrinos are Majorana particles, then only the ``chimney'' remains 
as allowed parameter space, where again the width is considerably 
reduced by future measurements of $\theta_{13}$.
The width of the ``chimney'' is also relevant for future improvements
of the cosmological mass bounds for neutrinos. The value of $\theta_{13}$
sets an upper bound for the sum of neutrino masses, which may be reached
by the cosmological bounds. This could lead to interesting scenarios
depending on whether \obb~experiments, cosmological determinations and/or 
improved $\theta_{13}$ measurements see a signal or improve the limits, 
respectively.
In the region of degenerate neutrino masses we found that improved
values of $\theta_{13}$ reduce the range of allowed masses on the 
lower side of $|m_{ee}|$.
Altogether we demonstrated, that there is a sizable interplay of 
the improvements expected in \obb~experiments, improved cosmological 
bounds and upcoming $\theta_{13}$ measurements.

\vspace{0.5cm}
{\bf Acknowledgments: }
This work has been supported by SFB375 (M.L.) and project 
number RO--2516/3--1 (W.R.) of Deutsche Forschungsgemeinschaft. 
We would like to thank M.~Goodman for discussions and 
J.~Kopp for help concerning computer problems.

\end{document}